\documentclass[a4paper]{article}

\usepackage[utf8]{inputenc}
\usepackage[top=10mm]{geometry}
\usepackage{booktabs}
\usepackage[flushleft]{threeparttable}
\usepackage{authblk}
\usepackage{bm}
\usepackage{amsmath}
\usepackage{amssymb}
\usepackage{mathrsfs}
\usepackage{graphicx}
\usepackage{indentfirst}
\usepackage[square,sort,numbers,compress]{natbib}
\usepackage{slashed}
\usepackage{float}
\usepackage{longtable,booktabs}
\usepackage{makecell}
\usepackage[figuresright]{rotating}
\usepackage[colorlinks,
linkcolor=blue,
anchorcolor=blue,
urlcolor=blue,
citecolor=blue
]{hyperref}
 \allowdisplaybreaks[4]

\title{\textbf{Octet baryon and heavy meson interactions in chiral effective field theory}}

\author[1,2]{Bo-Lin Huang \thanks{blhuang@imu.edu.cn}}
\author[3,4,5]{Bo Wang \thanks{wangbo@hbu.edu.cn}}
\author[6]{Shi-Lin Zhu \thanks{zhusl@pku.edu.cn}}

\affil[1]{\textit{\small School of Physical Science and Technology, Inner Mongolia University, Hohhot 010021, China}}
\affil[2]{\textit{\small Center for Quantum Physics and Technologies, School of Physical Science and Technology, Inner Mongolia University, Hohhot 010021, China}}
\affil[3]{\textit{\small College of Physics Science $\&$ Technology, Hebei University, Baoding 071002, China}}
\affil[4]{\textit{\small Hebei Key Laboratory of High-precision Computation and Application of Quantum Field Theory, Baoding, 071002, China}}
\affil[5]{\textit{\small Hebei Research Center of the Basic Discipline for Computational Physics, Baoding, 071002, China}}
\affil[6]{\textit{\small School of Physics and Center of High Energy
Physics, Peking University, Beijing 100871, China}}

\date{\small \today}

\begin{document}
\maketitle

\begin{abstract}
We calculate the effective potentials of the octet baryon and heavy meson systems using the chiral effective field theory up to the next-to-leading order. We consider the contact terms, one-pseudoscalar-meson and two-pseudoscalar-meson exchange contributions, facilitating a comprehensive analysis of the short-, mid-, and long-range interactions in these systems. The low energy constants are correlated with those of the $\bar{N}N$ interaction using a quark-level Lagrangian approach. We also incorporate the decuplet baryon contributions in the loop diagrams. Our research provides new insights into several near-threshold charmed baryons [e.g., $\Lambda_{c}(2940)$, $\Xi_{c}(3055)$, and $\Omega_{c}(3188)$, etc.] around $3$ GeV from the hadronic molecular perspective. We also observe several molecular states, referred to as $\Xi_{c}$, within the mass range of $3100-3500$ MeV. Further measurements of their resonance parameters and decay patterns in experiments may help to discriminate the conventional baryon and hadronic molecule explanations for these near-threshold states. 
\end{abstract}

\section{Introduction}

Investigations of the baryon-meson interactions provide significant insights into quantum chromodynamics (QCD) at hadronic scales, forming a critical foundation for advancing hadron spectroscopy. The quark model, known for its efficacy in demystifying hadron spectra \cite{ParticleDataGroup:2022pth}, faces substantial challenges in incorporating some near-threshold states such as $X(3872)$ \cite{Belle:2003nnu} and $D_{s0}(2317)$ \cite{BaBar:2003oey,Belle:2003guh,CLEO:2003ggt} within its theoretical framework \cite{Chen:2016qju,Guo:2017jvc,Liu:2019zoy,Lebed:2016hpi,Esposito:2016noz,Brambilla:2019esw,Meng:2022ozq}. These complications are similarly evident in charmed baryons, exemplified by $\Lambda_{c}(2940)$ and $\Xi_{c}(3055)$, which mirror the perplexities encountered with $X(3872)$ and $D_{s0}(2317)$.

The charmed baryon $\Lambda_{c}(2940)^+$ was first reported in 2007 by the BABAR Collaboration \cite{BaBar:2006itc}. It was identified in the $D^{0}p$ invariant mass spectrum. The $\Lambda_c(2940)^{+}$ was an isosinglet, evidenced by the absence of any signals in the $D^+p$ final state. This discovery was later corroborated by the Belle experiment, which uncovered the decay mode of  $\Lambda_{c}(2940)^{+} \rightarrow \Sigma_{c}^{0,++}\pi^{+,-}$ \cite{Belle:2006xni}. Further advancements were made in 2017 when the LHCb Collaboration shed light on the quantum numbers $J^P$ of $\Lambda_{c}(2940)^{+}$, suggesting $J^{P} = 3/2^-$ as the most likely spin-parity assignment \cite{LHCb:2017jym}.

There are two primary hypotheses on the internal structure of $\Lambda_{c}(2940)$: either a traditional charmed baryon or a $D^{*}N$ molecular state. The latter proposition arises primarily due to the challenge of categorizing $\Lambda_{c}(2940)$ as a $2P$ state in the charmed baryon spectroscopy. Its mass falls short by approximately $60-100$ MeV compared to the quark model predictions~\cite{Capstick:1986ter,Ebert:2011kk,Chen:2014nyo,Lu:2016ctt}. The molecular state hypothesis gained traction considering that $\Lambda_{c}(2940)$ is situated roughly $6$ MeV below the $D^{*0} p$ threshold. The authors of ref.~\cite{He:2006is} analyzed its decay patterns and suggested $\Lambda_{c}(2940)$ as a $1/2^{-}$ molecular state. He \textit{et al.} \cite{He:2010zq} examined the $D^{*} N$ interaction via the one-boson-exchange model and endorsed the view of $\Lambda_{c}(2940)$ as a $D^{*} N$ bound state with $I(J^P) = 0(1/2^+)$ or $0(3/2^-)$. Ortega \textit{et al.} ~\cite{Ortega:2012cx} delved into the $D^{*} N$ molecular framework using the constituent quark model and deduced a binding solution in the isoscalar $J^P = 3/2^{-}$ channel. This molecular picture is further supported by the analyses of the strong and radiative decays of $\Lambda_{c}(2940)$ in refs.~\cite{Dong:2009tg,Dong:2010xv}. Moreover, a QCD sum rule study ~\cite{Zhang:2012jk} posits that $\Lambda_{c}(2940)$ does not manifest as a compact state. This perspective gains additional backing from recent calculations within the chiral quark model ~\cite{Zhao:2016zhf}. For a more in-depth exploration and diverse viewpoints on the nature and characteristics of $\Lambda_{c}(2940)$, one might consult the comprehensive reviews and related studies in refs.~\cite{Klempt:2009pi,Crede:2013kia,Cheng:2015iom,Chen:2016spr,Kato:2018ijx,Huang:2016ygf,Wang:2015rda,Cheng:2015naa,Xie:2015zga,Romanets:2012hm,He:2011jp,Cheng:2006dk,Chen:2007xf,Zhong:2007gp,Chen:2015kpa,Luo:2019qkm,Yan:2022nxp,Luo:2022cun,Zhang:2022pxc,Yan:2023ttx,Xin:2023gkf,Ozdem:2023eyz,Yang:2023fsc,Wang:2023wii}.

Baryons composed of one up or down quark, one strange quark, and one charmed quark are classified as $\Xi_c$ baryons \cite{ParticleDataGroup:2022pth}. The landscape of charmed strange baryons has witnessed significant enrichment in recent years with the discovery of several highly excited states. The $\Xi_{c}(2980)$ and $\Xi_{c}(3080)$ baryons were first identified by the Belle Collaboration \cite{Belle:2006edu} and subsequently confirmed by the BABAR Collaboration \cite{BaBar:2007zjt}. Additionally, the BABAR Collaboration observed the $\Xi_{c}(3055)^{+}$ and $\Xi_{c}(3123)^{+}$ baryons \cite{BaBar:2007zjt}. However, only the $\Xi_{c}(3055)^{+}$ was later confirmed by the Belle Collaboration \cite{Belle:2013htj}. The Belle Collaboration \cite{Belle:2016tai} reported three charmed strange baryons. The $\Xi_{c}(3055)^{0}$ was first observed in the $\Lambda D^{0}$ decay mode, while the $\Xi_{c}(3055)^{+}$ and $\Xi_{c}(3080)^{+}$ were initially detected in the $\Lambda D^{+}$ decay mode. A notable aspect of these experiments is that the $J^P$ (total angular momentum and parity) quantum numbers for most of the observed charmed baryons have not been determined.

Since its discovery, the $\Xi_{c}(3055)$ baryon has been the subject of extensive study. Liu \textit{et al.} proposed that the $\Xi_{c}(3055)$ is a $D$-wave state based on their analysis of its strong decay \cite{Liu:2007ge}. Various studies have suggested different possibilities for the $J^P$ of the $\Xi_{c}(3055)$, including $3/2^+$, $5/2^+$, and $7/2^+$, as detailed in refs.~\cite{Guo:2008he, Liu:2012sj, Chen:2014nyo, Zhao:2016qmh, Chen:2016phw, Chen:2017aqm, Wang:2017vtv, Yao:2018jmc}. Additionally, Ye \textit{et al.} proposed in refs.~\cite{Ye:2017yvl, Ye:2017dra} that the $\Xi_{c}(3055)$ is not a $P$-wave excited $\Xi_{c}$ baryon, but rather a $2S$-wave state in the $^{3}P_{0}$ model. Contrary to these views, a molecular state description of the $\Xi_{c}(3055)$ was provided in ref.~\cite{Yu:2018yxl}, suggesting its $J^P$ as $1/2^-$. Given these diverse perspectives, it is clear that further research is essential to fully comprehend the nature of the $\Xi_{c}(3055)$ baryon.

The exploration of the octet baryon and heavy meson interactions is fundamental to resolving the mysteries associated with these charmed baryons. Moreover, a deep understanding of the octet baryon and heavy meson interactions is key to investigating the $D$-mesic nuclei \cite{Tsushima:1998ru,Garcia-Recio:2010fiq} as well as the characteristics of charmed mesons in nuclear environments \cite{Hosaka:2016ypm,Krein:2017usp}. An alternative approach, the meson-exchange model \cite{Machleidt:1987hj}, has been effectively utilized by the J\"{u}lich group \cite{Haidenbauer:2007jq,Haidenbauer:2008ff,Haidenbauer:2010ch} for the development of the $DN$ and $D^{*} N$ interaction framework.

The conceptual focus of modern nuclear force theory has shifted from the boson-exchange model to an approach more steeped in the foundational work of Weinberg \cite{Weinberg:1990rz,Weinberg:1991um} and mainly shaped within the framework of effective field theory. Particularly, the chiral effective field theory (ChEFT) has been widely and effectively utilized to examine the nucleon-nucleon ($NN$) interaction with significant success \cite{Bernard:1995dp, Epelbaum:2008ga, Machleidt:2011zz, Meissner:2015wva, Hammer:2019poc,Machleidt:2020vzm}. This theory has also shown considerable promise in the investigation of systems incorporating heavy flavors, as exemplified in various studies \cite{Liu:2012vd,Meng:2019ilv,Wang:2019ato,Meng:2019nzy,Wang:2019nvm,Wang:2020dko,Cheng:2023vyv,Lin:2023ihj} (see ref.~\cite{Meng:2022ozq} for a recent review). The utility of the approach becomes apparent in its diverse application: deciphering the newly discovered pentaquarks \cite{Wang:2019ato}, and extrapolating the $\Sigma_{c}N$ potential from the lattice QCD data to the physical pion mass \cite{Meng:2019nzy} being some prominent examples. 

In ref.~\cite{Wang:2020dhf}, we examined the $D^{(*)}N$ interactions through ChEFT. We extend our earlier research in this work. The findings from refs.~\cite{Kaiser:2001hr,Liu:2006xja,Huang:2015ghe,Huang:2017bmx,Huang:2019not,Huang:2020iai,Huang:2021ulf,Huang:2021fdt,Lu:2021gsb,Lu:2022hwm,Huang:2022cag} suggest that employing the SU(3) framework for such calculations yields plausible predictions. This current study aims to further explore the interactions between octet baryons and heavy mesons up to next-to-leading order within the ChEFT. Our approach is comprehensive, incorporating the long-, mid-, and short-range interactions. In the conventional meson exchange models, the short-range interaction is typically described by the exchange of heavy mesons, with the $\omega(782)$ meson being one notable example. However, ChEFT takes a different approach by considering an expansion in small momenta $Q$, where $Q$ is too small to resolve structures such as the $\rho(770)$ or $\omega(782)$ mesons, because $Q\ll \Lambda_{\chi} \approx m_{\rho,\omega}$. Thus, when building an effective theory like ChEFT, Lepage suggested the following three steps \cite{Lepage:1997cs}:
\begin{enumerate}
\item Incorporate the correct long-range behavior: The long-range behavior of the underlying theory must be known, and it must be built into the effective theory. In the case of the octet baryon and heavy meson systems, the long-range interactions, encompassing both long-range and mid-range scales, are mainly described by the one and two pseudoscalar meson exchanges.
\item Introduce an ultraviolet cutoff to exclude the high-momentum states, or, equivalent, to soften the short-distance behavior.
The cutoff has two effects: (i) it excludes high-momentum states, which are sensitive to the unknown short-distance dynamics, and only states that we understand are retained. (ii) it makes all interactions regular at $r = 0$, thereby avoiding the infinities.
\item Add local correction terms (also known as contact or counterterms) to the effective Hamiltonian. These mimic the effects of the high-momentum states excluded by the cutoff introduced in the previous step. In the conventional meson-exchange picture, the short-range interaction is described by heavy meson exchange, like the $\rho(770)$ and $\omega(782)$. However, at low energy,
such structures are not resolved. Since we must include contact terms anyhow, it is most efficient to use them to account for any heavy meson exchange as well. The correction terms systematically remove dependence on the cutoff.
\end{enumerate}

Additionally, we include the contributions of the decuplet baryons, considering them as intermediate states in the loops. Utilizing ChEFT, we not only calculate the effective potentials between the octet baryons and heavy mesons but also investigate the possible bound states.

This paper is structured as follows. Section~\ref{lagrangian} introduces the chiral Lagrangians. Section~\ref{potentials} covers Feynman diagrams and expressions of the effective potentials. Section~\ref{results} presents and discusses our findings. The final section provides a brief summary. The parameters for the potentials are detailed in Appendix~\ref{AppParameters}. Appendix~\ref{AppDecayC} and \ref{AppRegulator} delve into the impact of the decay constant and the regulator function, respectively.

\section{Chiral Lagrangian}
\label{lagrangian}  
Our calculation of the octet baryon and heavy meson interaction is based on the SU(3) effective chiral Lagrangian in the heavy hadron formulation,
\begin{align}
\label{eq1}
\mathcal{L}_{\text{eff}}=\mathcal{L}_{B\phi}+\mathcal{L}_{
B \phi  T}+\mathcal{L}_{H
\phi}+\mathcal{L}_{BH}.
\end{align}
The traceless hermitian $3\times 3$ matrices $B$ and $\phi$ include the octet baryon fields ($N, \Lambda, \Sigma, \Xi$) and the light pseudoscalar Goldstone boson fields ($\phi=\pi,K,\bar{K},\eta$), respectively. The $T$ represents the decuplet baryon fields ($\Delta,\Sigma^*,\Xi^*,\Omega$). The $H$ is the superfield for the charmed mesons. The lowest-order SU(3) chiral Lagrangian for the octet baryon and light meson interaction take the form \cite{Borasoy:1996bx}
\begin{align}
\label{eq2}
 \mathcal{L}_{B \phi }^{(1)}&=\text{tr}(i\bar{B}[v\cdot D,B])+2 D\text{tr}(\bar{B}S_{\mu}\{u^{\mu},B\})+2 F\text{tr}(\bar{B}S_{\mu}[u^{\mu},B]),
\end{align}
where $\text{tr}(\cdots)$ represents the trace in flavor space, $v_\mu=(1,0,0,0)$ is the heavy meson velocity, $D_{\mu}$ denotes the covariant derivative
\begin{align}
\label{eq3}
[D_{\mu},B]=\partial_{\mu}B+[\Gamma_{\mu},B],
\end{align}
and $S_{\mu}$ is the covariant spin operator. In practice one works with $\vec{\sigma}$ (the Pauli spin matrices)
\begin{align}
\label{eq4}
S^{\mu}=(0,\frac{\vec{\sigma}}{2}).
\end{align}
The chiral connection $\Gamma^{\mu}=[\xi^{\dag},\partial^{\mu}\xi]/2$ and the axial vector quantity $u^{\mu}=i\{\xi^{\dag},\partial^{\mu}\xi\}/2$ contain an even and odd number of meson fields, respectively. The SU(3) matrix $U=\xi^{2}=\text{exp}(i\phi/f)$ collects the light pseudoscalar Goldstone boson fields. The parameter $f$ is the pseudoscalar decay constant in the chiral limit. The axial vector coupling constants $D$ and $F$ can be determined by fitting the semi-leptonic decays of the octet baryons \cite{Borasoy:1998pe}. The lowest-order SU(3) chiral Lagrangian involving the decuplet baryon takes the form \cite{Lebed:1993yu}

\begin{align}
\label{eq5}
\mathcal{L}_{
B \phi T}^{(1)}=-\bar{T}^{\mu}(i v\cdot D-\delta_B)T_\mu+\mathcal{C}(\bar{T}^\mu u_\mu B+\bar{B}u_\mu T^{\mu})+2\mathcal{H}\bar{T}^{\mu}S\cdot u T_\mu,
\end{align}
where $\delta_B$ denotes the mass difference between the decuplet and octet baryon in the chiral limit. The chiral covariant derivative with the decuplet baryon field reads
\begin{align}
\label{eq6}
iD_\mu{T_{abc}^\nu}=i\partial_\mu{T_{abc}^\nu}+(\Gamma_\mu)_a^d{T_{dbc}^\nu}+(\Gamma_\mu)_b^d{T_{adc}^\nu}+(\Gamma_\mu)_c^d{T_{abd}^\nu},
\end{align}
with 
\begin{align}
\label{eq7}
T_{abc}=\left(
\begin{array}{ccc}
\Delta^{++} & \frac{\Delta^{+}}{\sqrt{3}} & \frac{\Sigma^{*+}}{\sqrt{3}} \\\\
\frac{\Delta^{+}}{\sqrt{3}} & \frac{\Delta^{0}}{\sqrt{3}} & \frac{\Sigma^{*0}}{\sqrt{6}} \\\\
\frac{\Sigma^{*+}}{\sqrt{3}} & \frac{\Sigma^{*0}}{\sqrt{6}} & \frac{\Xi^{*0}}{\sqrt{3}} \end{array}\right)\left(
\begin{array}{ccc}
\frac{\Delta^{+}}{\sqrt{3}} & \frac{\Delta^{0}}{\sqrt{3}} & \frac{\Sigma^{*0}}{\sqrt{6}} \\\\
\frac{\Delta^{0}}{\sqrt{3}}  & \Delta^{-}  & \frac{\Sigma^{*-}}{\sqrt{3}} \\\\
\frac{\Sigma^{*0}}{\sqrt{6}} & \frac{\Sigma^{*-}}{\sqrt{3}} & \frac{\Xi^{*-}}{\sqrt{3}} \end{array}\right)\left(
\begin{array}{ccc}
\frac{\Sigma^{*+}}{\sqrt{3}}  & \frac{\Sigma^{*0}}{\sqrt{6}}  & \frac{\Xi^{*0}}{\sqrt{3}} \\\\
\frac{\Sigma^{*0}}{\sqrt{6}}  & \frac{\Sigma^{*-}}{\sqrt{3}}  & \frac{\Xi^{*-}}{\sqrt{3}} \\\\
\frac{\Xi^{*0}}{\sqrt{3}} & \frac{\Xi^{*-}}{\sqrt{3}} & \Omega^{-} \end{array}\right).
\end{align}
In this representation, one can assign any particular permutation of indices $a$, $b$, $c$ to denote the row, column, and sub-matrix. The decuplet remains completely symmetric after rearranging the flavor indices.

The lowest-order chiral Lagrangian for the heavy mesons reads \cite{Wise:1992hn,Manohar:2000dt}
\begin{align}
\label{eq8}
 \mathcal{L}_{H \phi}^{(1)}=-\left\langle(iv\cdot\partial H)\bar{H}\right\rangle+\left\langle H(i v\cdot \Gamma)\bar{H}\right\rangle-\frac{1}{8}\delta_D\left\langle H \sigma^{\mu\nu}\bar{H}\sigma_{\mu\nu}\right\rangle+g\left\langle H u_\mu \gamma^\mu \gamma_5 \bar{H}\right\rangle,
\end{align}
where $\left\langle \cdots \right\rangle$ represents the trace in spinor space and $\delta_D$ denotes the mass difference between the heavy pseudoscalar and vector meson. The doublet of the ground state heavy mesons reads
\begin{align}
\label{eq9}
H=\frac{1+\slashed{v}}{2}(P_\mu^{*}\gamma^{\mu}+iP\gamma_5),\quad
\bar{H}=\gamma^{0}H^{\dag}\gamma^{0}=(P_\mu^{*\dag}\gamma^{\mu}+iP^{\dag}\gamma_5)\frac{1+\slashed{v}}{2},
\end{align}
\begin{align}
\label{eq10}
P=(D^{0},D^{+},D_s^{+}),\quad
P_\mu^{*}=(D^{0*},D^{+*},D_{s}^{+*})_\mu.
\end{align}

At last, we construct the leading order (LO) and the next-to-leading order (NLO) contact Lagrangians between the octet baryon and heavy meson. Because of parity, the contact interactions come only in even powers of derivatives or masses, thus,
\begin{align}
\label{eq11} 
\mathcal{L}_{BH}^{(0)}=&D_a \text{tr}(\bar{B}B) \left\langle  H \bar{H} \right\rangle+D_b \text{tr}(\bar{B} \gamma_\mu \gamma_5 B) \left\langle H \gamma^\mu \gamma_5 \bar{H}\right\rangle\nonumber\\
&+E_a \text{tr}(\bar{B}\lambda_a B) \left\langle  H  \lambda_a \bar{H} \right\rangle+E_b \text{tr}(\bar{B} \gamma_\mu \gamma_5 \lambda_a B) \left\langle H \gamma^\mu \gamma_5 \lambda_a \bar{H}\right\rangle,
\end{align}
\begin{align}
\label{eq11} 
\mathcal{L}_{BH}^{(2,h)}=&D_a^{h} \text{tr}(\bar{B}B) \left\langle  H \bar{H} \right\rangle\text{tr}(\chi_{+})+D_b^{h} \text{tr}(\bar{B} \gamma_\mu \gamma_5 B) \left\langle H \gamma^\mu \gamma_5 \bar{H}\right\rangle\text{tr}(\chi_{+})\nonumber\\
&+E_a^{h} \text{tr}(\bar{B}\lambda_a B) \left\langle  H  \lambda_a \bar{H} \right\rangle\text{tr}(\chi_{+})+E_b^{h} \text{tr}(\bar{B} \gamma_\mu \gamma_5 \lambda_a B) \left\langle H \gamma^\mu \gamma_5 \lambda_a \bar{H}\right\rangle\text{tr}(\chi_{+}),
\end{align}
\begin{align}
\label{eq11} 
\mathcal{L}_{BH}^{(2,v)}=&\{D_{a1}^{v} \text{tr}[(v\cdot D\bar{B})(v\cdot D B)] \left\langle  H \bar{H} \right\rangle+D_{a2}^{v} \text{tr}[((v\cdot D)^2\bar{B})  B ] \left\langle  H \bar{H} \right\rangle\nonumber\\
&+D_{a3}^{v} \text{tr}(\bar{B} B) \left\langle  (v\cdot D H) (v\cdot D \bar{H}) \right\rangle+ D_{a4}^{v} \text{tr}(\bar{B} B) \left\langle  ((v\cdot D)^2 H) \bar{H} \right\rangle\nonumber\\
&+D_{a5}^{v} \text{tr}[(v\cdot D\bar{B}) B ] \left\langle  (v\cdot D H) \bar{H} \right\rangle+D_{a6}^{v} \text{tr}[(v\cdot D\bar{B}) B ] \left\langle  H (v\cdot D \bar{H}) \right\rangle\nonumber\\
&+D_{b1}^{v} \text{tr}[(v\cdot D\bar{B}) \gamma_\mu \gamma_5 (v\cdot D B)] \left\langle H \gamma^\mu \gamma_5 \bar{H}\right\rangle+\cdots+E_{a1}^{v} \text{tr}[(v\cdot D\bar{B})\lambda_a (v\cdot D B)] \left\langle  H  \lambda_a \bar{H} \right\rangle\nonumber\\
&+\cdots+E_{b1}^{v} \text{tr}[(v \cdot D \bar{B}) \gamma_\mu \gamma_5 \lambda_a (v\cdot D B)] \left\langle H \gamma^\mu \gamma_5 \lambda_a \bar{H}\right\rangle+\cdots\}+\text{H.c.},
\end{align}
\begin{align}
\label{eq11} 
\mathcal{L}_{BH}^{(2,q)}=&\{D_{1}^{q} \text{tr}[(D^\mu \bar{B}) \gamma_\mu \gamma_5 (D^\nu B)] \left\langle H \gamma_\nu \gamma_5 \bar{H}\right\rangle+D_{2}^{q} \text{tr}[(D^\mu D^\nu \bar{B}) \gamma_\mu \gamma_5 B] \left\langle H \gamma_\nu \gamma_5 \bar{H}\right\rangle \nonumber\\
&+D_{3}^{q} \text{tr}(\bar{B} \gamma_\mu \gamma_5  B) \left\langle (D^\mu H) \gamma_\nu \gamma_5 (D^\nu \bar{H})\right\rangle+D_{4}^{q} \text{tr}(\bar{B} \gamma_\mu \gamma_5  B) \left\langle (D^\mu D^\nu H) \gamma_\nu \gamma_5 \bar{H} \right\rangle\nonumber\\
&+D_{5}^{q} \text{tr}[(D^\mu \bar{B}) \gamma_\mu \gamma_5  B] \left\langle (D^\nu H) \gamma_\nu \gamma_5 \bar{H} \right\rangle+D_{6}^{q} \text{tr}[(D^\mu \bar{B}) \gamma_\mu \gamma_5  B] \left\langle  H \gamma_\nu \gamma_5 (D^\nu \bar{H}) \right\rangle\nonumber\\
&+E_{1}^{q} \text{tr}[(D^\mu \bar{B}) \gamma_\mu \gamma_5 \lambda_a (D^\nu B)] \left\langle H \gamma_\nu \gamma_5 \lambda_a \bar{H}\right\rangle+\cdots\}+\text{H.c.},\cdots,
\end{align}
where $D_a$ and $D_b$ are two low energy constants (LECs) that contribute to the central potential and the spin-spin interaction, respectively. On the other hand, $E_a$ and $E_b$ are associated with the isospin-isospin interaction, influencing the central potential and spin-spin interaction, respectively. The $\lambda_a$ represents the conventional Gell-Mann matrix. The quantity $\chi_{+}=\xi^\dag\chi\xi^\dag+\xi\chi\xi$ with $\chi=\text{diag}(m_{\pi}^2,m_\pi^2,2m_K^2-m_\pi^2)$ introduces explicit chiral symmetry breaking terms. These contact terms are needed to parametrize the unresolved short-distance dynamics of the octet baryon and heavy meson systems, to make results reasonably independent of regulators, and to renormalize loop integrals.

\section{Expressions of the effective potentials}
\label{potentials}
\begin{figure}[t]
\centering
\includegraphics[height=2.5cm,width=5.6cm]{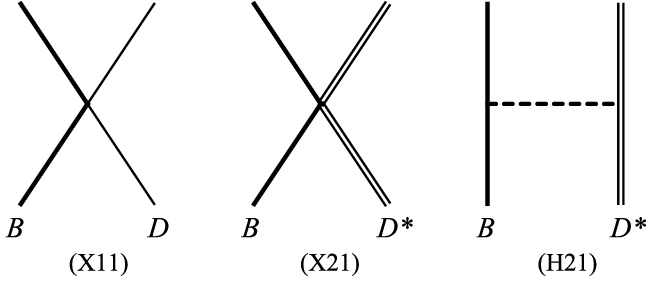}
\caption{\label{fig:BDandBDstartree}The leading order Feynman diagrams $\mathcal{O}(\epsilon^0)$ in the calculation of the octet baryon ($B$) and the heavy meson ($D/D^*$) effective potentials. The thick, thin, double-thin, and dashed lines denote the $B$, $D$, $D^*$, and $\phi$ fields, respectively.}
\end{figure}

\begin{figure}[t]
\centering
\includegraphics[height=5cm,width=7.0cm]{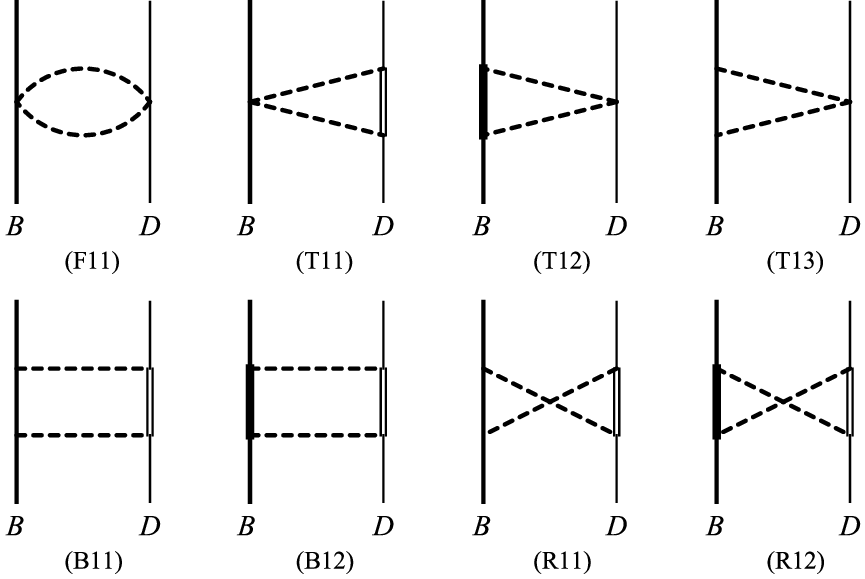}
\caption{\label{fig:BDoneloop}The two-pseudoscalar-meson-exchange diagrams for the $BD$ system at $\mathcal{O}(\epsilon^2)$. These diagrams are categorized into the football diagram (F11), triangle diagrams (T11,T12,T13), box diagrams (B11,B12), and crossed box diagrams (R11,R12). The decuplet fields are represented by heavy-thick lines within the loops. The notations for other elements remain consistent with those presented in Fig.~\ref{fig:BDandBDstartree}.}
\end{figure}

\begin{figure}[t]
\centering
\includegraphics[height=10cm,width=7.0cm]{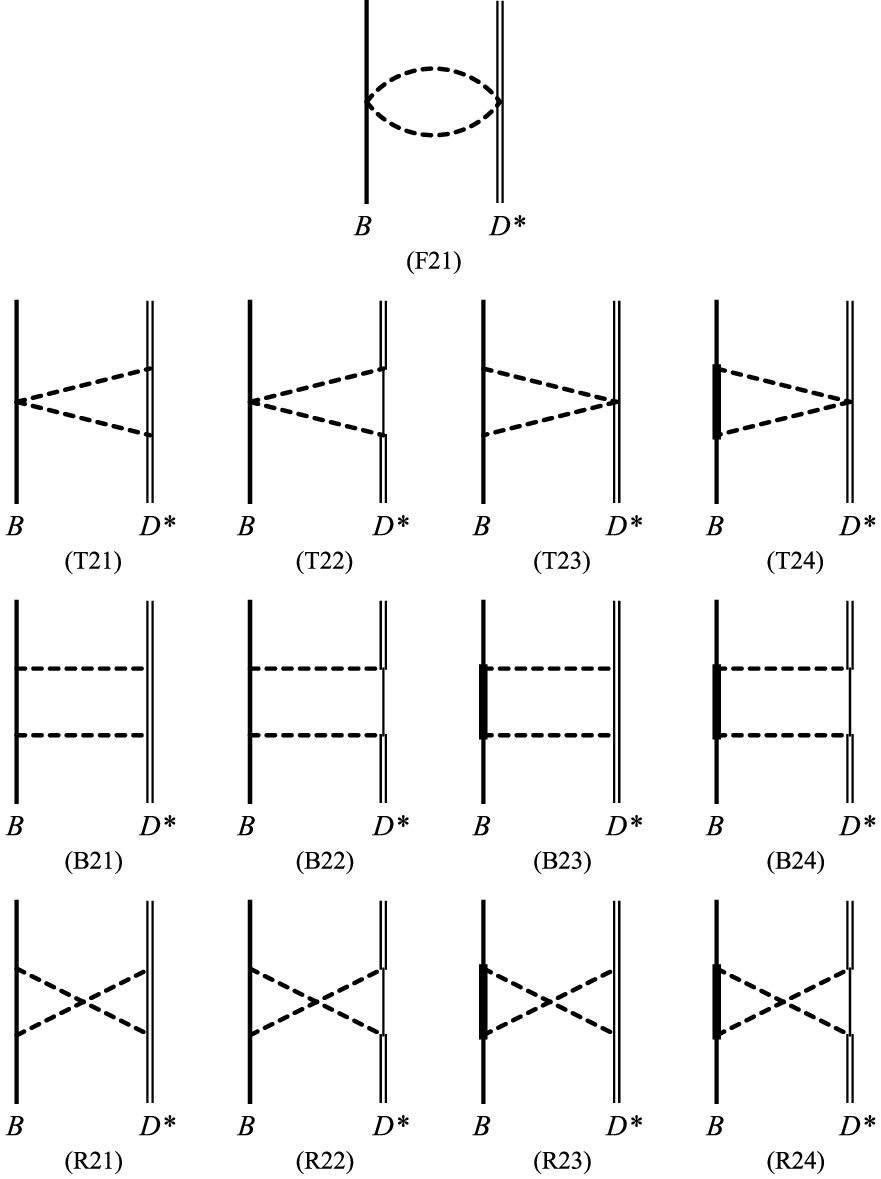}
\caption{\label{fig:BDstaroneloop}The two-pseudoscalar-meson-exchange diagrams for the $BD^*$ system at $\mathcal{O}(\epsilon^2)$. The notations remain consistent with those depicted in Fig.~\ref{fig:BDoneloop}.}
\end{figure}

\begin{figure}[t]
\centering
\includegraphics[height=10cm,width=7.0cm]{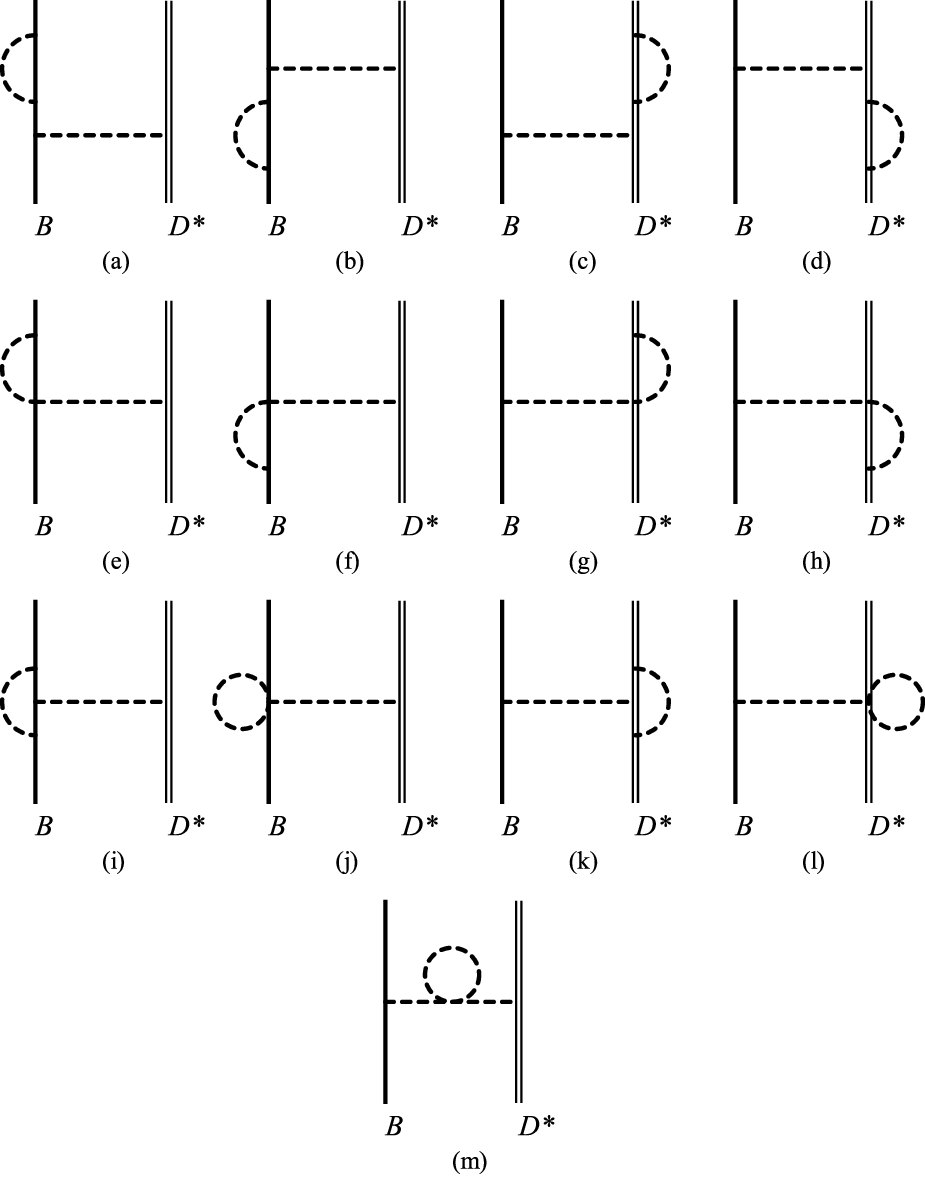}
\caption{\label{fig:BDstarOneMesonOneloop} Diagrams contributing to the renormalization of the one-pseudoscalar-meson exchange. The notations remain consistent with those depicted in Fig.~\ref{fig:BDandBDstartree}.}
\end{figure}

In the context of heavy hadron chiral perturbation theory, the scattering amplitudes for the $BD$ and $BD^*$ systems can be systematically expanded in a series. These expansions are based on small dimensionless parameters, denoted as $\epsilon=\{q/\Lambda_\chi,m_\phi/\Lambda_\chi,\delta/\Lambda_\chi\}$. Here, $\Lambda_\chi\simeq 1\,\text{GeV}$ represents the scale of chiral symmetry breaking. The parameter $q$ refers to the momentum of Goldstone bosons or the residual momentum of heavy hadrons, $m_\phi$ indicates the masses of the light pseudoscalar mesons, and $\delta$ signifies the mass gap between the decuplet and octet baryons or between the heavy pseudoscalar and vector mesons. This power counting scheme is often denoted as the small scale expansion (SSE). Following the power counting of nucleon-nucleon interaction \cite{Machleidt:2011zz}, the power formula read
\begin{align}
   \epsilon=2L+\sum_{i} \Delta_i, 
\end{align}
with
\begin{align}
  \Delta_i= d_i+\frac{n_i}{2}-2,
\end{align}
where $L$ denotes the number of loops in the diagram; $d_i$ represents the number of derivatives or small-mass insertions, and $n_i$ denotes the number of heavy fields involved in vertex $i$. The sum is taken over all vertices $i$ within the diagram under consideration. At leading order (LO, $\epsilon=0$), the potentials consist of $\mathcal{O}(\epsilon^0)$ contact and one-meson exchange terms, represented by the (X11, X21) and (H21) diagrams  shown in Fig.~\ref{fig:BDandBDstartree}. In the next order, $\epsilon=1$, all contributions vanish due to parity and time-reversal invariance. Therefore, the next-to-leading order (NLO) is $\epsilon=2$. Two-meson-exchange occurs at this order, see Figs.~\ref{fig:BDoneloop}-\ref{fig:BDstaroneloop}. Furthermore, there are some $\mathcal{O}(\epsilon^2)$ contact terms which originate from diagrams similar to (X11, X21) of Fig.~\ref{fig:BDandBDstartree}, but with vertices corresponding to the NLO contact Lagrangians.

We present the explicit expressions of the effective potentials for the $BD$ and $BD^*$ systems at $\mathcal{O}(\epsilon^0)$. The corresponding Feynman diagrams are shown in Fig.~\ref{fig:BDandBDstartree}, which contain the contact and one-pseudoscalar-meson exchange diagrams. The one-pseudoscalar-meson exchange diagram for the $BD$ system vanishes because the three pseudoscalar meson vertex is forbidden. Then, the potentials read
\begin{align}
\label{eq12}
\mathcal{V}_{BD}^{\text{(X11)}}=D_a+\alpha^{\text{(X11)}}E_a,
\end{align}
\begin{align}
\label{eq13}
\mathcal{V}_{BD^*}^{\text{(X21)}}=D_a+D_b\bm{\sigma}\cdot\bm{T}+\alpha^{\text{(X21)}}(E_a+E_b\bm{\sigma}\cdot\bm{T}),
\end{align}
\begin{align}
\label{eq14}
\mathcal{V}_{BD^*}^{\text{(H21)}}=\alpha^{\text{(H21)}}_\pi \frac{g}{f^2}\frac{(\bm{\sigma}\cdot\bm{q})(\bm{T}\cdot\bm{q})}{\bm{q}^2+m_\pi^2}+\alpha^{\text{(H21)}}_\eta \frac{g}{f^2}\frac{(\bm{\sigma}\cdot\bm{q})(\bm{T}\cdot\bm{q})}{\bm{q}^2+m_\eta^2},
\end{align}
where the parameters $\alpha^{\text{(X11)}}$, $\alpha^{\text{(X21)}}$, $\alpha^{\text{(H21)}}_\pi$, and $\alpha^{\text{(H21)}}_\eta$, which incorporate crucial isospin factors, are detailed in the Appendix~\ref{AppParameters}. The operators $\bm{\sigma}$ and $\bm{T}$ 
are connected to the spin operators of the spin-$\frac{1}{2}$ baryon and spin-$1$ meson, being represented as ($\frac{1}{2} \bm{\sigma}$) and ($-\bm{T}$) respectively. This relationship is elaborated in detail in ref.~\cite{Wang:2019ato}. The Breit approximation formula,
\begin{align}
\label{eq15}
\mathcal{V}(\bm{q})=-\frac{\mathcal{M}(\bm{q})}{\sqrt{2M_12M_22M_32M_4}}
\end{align}
is employed to establish a connection between the scattering amplitude $\mathcal{M}(\bm{q})$ and the effective potential $\mathcal{V}(\bm{q})$ in momentum space, where $M_{1,...,4}$ are the masses of the scattering particles.

The two-pseudoscalar-meson-exchange diagrams for the $BD$ system at order $\mathcal{O}(\epsilon^2)$ are shown in Fig.~\ref{fig:BDoneloop}. The effective potentials from these diagrams read
\begin{align}
\label{eq16}
\mathcal{V}_{BD}^{\text{(F11)}}=\alpha^{\text{(F11)}}_{\pi\pi}\frac{1}{f^4}J_{22}^{F}(m_\pi,m_\pi)+\alpha^{\text{(F11)}}_{KK}\frac{1}{f^4}J_{22}^{F}(m_K,m_K),
\end{align}
\begin{align}
\label{eq17}
\mathcal{V}_{BD}^{\text{(T11)}}=&\Big\{\frac{\alpha_{\pi\pi}^{\text{(T11)}}}{f^4},\frac{\alpha_{KK}^{\text{(T11)}}}{f^4}\Big\}g^2\Big[(d-1)J_{34}^{T}-(J_{24}^{T}+J_{33}^{T})\bm{q}^2\Big]\{(m_\pi,m_\pi,-\delta_D),(m_K,m_K,-\delta_D)\},
\end{align}
\begin{align}
\label{eq18}
\mathcal{V}_{BD}^{\text{(T12)}}=&\Big\{\frac{\alpha_{\pi\pi}^{\text{(T12)}}}{f^4},\frac{\alpha_{KK}^{\text{(T12)}}}{f^4}\Big\}\mathcal{C}^2\Big[(2-d)J_{34}^{T}-\frac{2-d}{d-1}(J_{24}^{T}+J_{33}^{T})\bm{q}^2\Big]\{(m_\pi,m_\pi,-\delta_B),(m_K,m_K,-\delta_B)\},
\end{align}
\begin{align}
\label{eq19}
\mathcal{V}_{BD}^{\text{(T13)}}=&\Big\{\frac{\alpha_{\pi\pi}^{\text{(T13)}}}{f^4},\frac{\alpha_{KK}^{\text{(T13)}}}{f^4}\Big\}\Big[(d-1)J_{34}^{T}-(J_{24}^{T}+J_{33}^{T})\bm{q}^2\Big]\{(m_\pi,m_\pi),(m_K,m_K)\},
\end{align}
\begin{align}
\label{eq20}
\mathcal{V}_{BD}^{\text{(B11)}}=&\Big\{\frac{\alpha_{\pi\pi}^{\text{(B11)}}}{f^4},\frac{\alpha_{KK}^{\text{(B11)}}}{f^4},\frac{\alpha_{\eta\eta}^{\text{(B11)}}}{f^4},\frac{\alpha_{\pi\eta}^{\text{(B11)}}}{f^4},\frac{\alpha_{\eta\pi}^{\text{(B11)}}}{f^4}\Big\}g^2\Big\{(d^2-1)J_{41}^{B}-[J_{21}^{B}+2(d+1)(J_{31}^{B}+J_{42}^{B})]\bm{q}^2\nonumber\\
&+(J_{22}^{B}+2J_{32}^{B}+J_{43}^{B})\bm{q}^4\Big\}\{(m_\pi,m_\pi,-\delta_D),(m_K,m_K,-\delta_D),(m_\eta,m_\eta,-\delta_D),\nonumber\\
&(m_\pi,m_\eta,-\delta_D),(m_\eta,m_\pi,-\delta_D)\},
\end{align}
\begin{align}
\label{eq21}
\mathcal{V}_{BD}^{\text{(B12)}}=&\Big\{\frac{\alpha_{\pi\pi}^{\text{(B12)}}}{f^4},\frac{\alpha_{KK}^{\text{(B12)}}}{f^4},\frac{\alpha_{\eta\eta}^{\text{(B12)}}}{f^4},\frac{\alpha_{\pi\eta}^{\text{(B12)}}}{f^4},\frac{\alpha_{\eta\pi}^{\text{(B12)}}}{f^4}\Big\}\mathcal{C}^2g^2\frac{d-2}{d-1}\Big\{(J_{22}^{B}+2J_{32}^{B}+J_{43}^{B})\bm{q}^4+(d^2-1)J_{41}^{B}\nonumber\\
&-[J_{21}^{B}+2(d+1)(J_{31}^{B}+J_{42}^{B})]\bm{q}^2\Big\}\{(m_\pi,m_\pi,-\delta_B,-\delta_D),(m_K,m_K,-\delta_B,-\delta_D),\nonumber\\
&(m_\eta,m_\eta,-\delta_B,-\delta_D),(m_\pi,m_\eta,-\delta_B,-\delta_D),(m_\eta,m_\pi,-\delta_B,-\delta_D)\},
\end{align}
\begin{align}
\label{eq22}
\mathcal{V}_{BD}^{\text{(R11)}}=\mathcal{V}_{BD}^{\text{(B11)}}|_{J_{ij}^{B}\rightarrow J_{ij}^{R},\alpha^{\text{(B11)}}_{**}\rightarrow\alpha^{\text{(R11)}}_{**}},
\end{align}
\begin{align}
\label{eq23}
\mathcal{V}_{BD}^{\text{(R12)}}=\mathcal{V}_{BD}^{\text{(B12)}}|_{J_{ij}^{B}\rightarrow J_{ij}^{R},\alpha^{\text{(B12)}}_{**}\rightarrow\alpha^{\text{(R12)}}_{**}},
\end{align}
where the parameters $\alpha^{(**)}_{**}$ are detailed in the Appendix~\ref{AppParameters}. The loop functions $J_{ij}^{F}$, $J_{ij}^{T}$, $J_{ij}^{B}$, and $J_{ij}^{R}$ are given in refs.~\cite{Meng:2019ilv,Wang:2019ato}. Nevertheless, due to the varying masses of the internal pseudoscalar meson fields in the SU(3) framework, we have employed the formulas $\bar{\Delta}=m_1^2+(m_2^2-m_1^2) x+q^2x(x-1)-i\epsilon$ and $A=m_1^2-\omega^2+(m_2^2-m_1^2) x+q^2x(x-1)-i\epsilon$ (where $m_1$ and $m_2$ denote the masses of the internal pseudoscalar meson fields), as substitutes for the equivalent expressions outlined in ref.~\cite{Wang:2019ato}. In addition, we take the scale $\lambda=4\pi f_\pi$ in the loop functions. The parameter $d$ denotes the dimension introduced in the dimensional regularization. The physical values of the pseudoscalar-meson decay constants have been incorporated into the analysis at next-to-leading order. Differences become noticeable at higher orders of calculation. Note that, to generate the effective potential, one must subtract the two-particle reducible contributions from the box diagrams. A detailed deduction, based on the principal value integral method, is provided in Appendix B of ref.~\cite{Wang:2019ato}.

The two-pseudoscalar-meson-exchange diagrams for the $BD^{*}$ system at order $\mathcal{O}(\epsilon^2)$ are depicted in Fig.~\ref{fig:BDstaroneloop}. The effective potentials arising from these diagrams are expressed as follows:
\begin{align}
\label{eq24}
\mathcal{V}_{BD^{*}}^{\text{(F21)}}=\alpha^{\text{(F21)}}_{\pi\pi}\frac{1}{f^4}J_{22}^{F}(m_\pi,m_\pi)+\alpha^{\text{(F21)}}_{KK}\frac{1}{f^4}J_{22}^{F}(m_K,m_K),
\end{align}
\begin{align}
\label{eq25}
\mathcal{V}_{BD^{*}}^{\text{(T21)}}=&\Big\{\frac{\alpha_{\pi\pi}^{\text{(T21)}}}{f^4},\frac{\alpha_{KK}^{\text{(T21)}}}{f^4}\Big\}g^2\frac{(d-3)(d-2)}{d-1}\Big[(d-1)J_{34}^{T}-(J_{24}^{T}+J_{33}^{T})\bm{q}^2\Big]\{(m_\pi,m_\pi),(m_K,m_K)\},
\end{align}
\begin{align}
\label{eq26}
\mathcal{V}_{BD^{*}}^{\text{(T22)}}=&\Big\{\frac{\alpha_{\pi\pi}^{\text{(T22)}}}{f^4},\frac{\alpha_{KK}^{\text{(T22)}}}{f^4}\Big\}g^2\Big[J_{34}^{T}-\frac{1}{d-1}(J_{24}^{T}+J_{33}^{T})\bm{q}^2\Big]\{(m_\pi,m_\pi,\delta_D),(m_K,m_K,\delta_D)\},
\end{align}
\begin{align}
\label{eq27}
\mathcal{V}_{BD^{*}}^{\text{(T23)}}=&\Big\{\frac{\alpha_{\pi\pi}^{\text{(T23)}}}{f^4},\frac{\alpha_{KK}^{\text{(T23)}}}{f^4}\Big\}\Big[(d-1)J_{34}^{T}-(J_{24}^{T}+J_{33}^{T})\bm{q}^2\Big]\{(m_\pi,m_\pi),(m_K,m_K)\},
\end{align}
\begin{align}
\label{eq28}
\mathcal{V}_{BD^{*}}^{\text{(T24)}}=&\Big\{\frac{\alpha_{\pi\pi}^{\text{(T24)}}}{f^4},\frac{\alpha_{KK}^{\text{(T24)}}}{f^4}\Big\}\Big[(2-d)J_{34}^{T}-\frac{2-d}{d-1}(J_{24}^{T}+J_{33}^{T})\bm{q}^2\Big]\{(m_\pi,m_\pi,-\delta_B),(m_K,m_K,-\delta_B)\},
\end{align}
\begin{align}
\label{eq29}
\mathcal{V}_{BD^{*}}^{\text{(B21)}}=&\Big\{\frac{\alpha_{\pi\pi}^{\text{(B21)}}}{f^4},\frac{\alpha_{KK}^{\text{(B21)}}}{f^4},\frac{\alpha_{\eta\eta}^{\text{(B21)}}}{f^4},\frac{\alpha_{\pi\eta}^{\text{(B21)}}}{f^4},\frac{\alpha_{\eta\pi}^{\text{(B21)}}}{f^4}\Big\}g^2\frac{(d-3)(d-2)}{d-1}\Big\{(d^2-1)J_{41}^{B}\nonumber\\
&-\Big[\Big(\frac{1}{d-2}\bm{\sigma}\cdot\bm{T}+1\Big)J_{21}^{B}+2(d+1)(J_{31}^{B}+J_{42}^{B})\Big]\bm{q}^2+(J_{22}^{B}+2J_{32}^{B}+J_{43}^{B})\bm{q}^4\Big\}\{(m_\pi,m_\pi),\nonumber\\
&(m_K,m_K),(m_\eta,m_\eta),(m_\pi,m_\eta),(m_\eta,m_\pi)\},
\end{align}
\begin{align}
\label{eq30}
\mathcal{V}_{BD^{*}}^{\text{(B22)}}=&\Big\{\frac{\alpha_{\pi\pi}^{\text{(B22)}}}{f^4},\frac{\alpha_{KK}^{\text{(B22)}}}{f^4},\frac{\alpha_{\eta\eta}^{\text{(B22)}}}{f^4},\frac{\alpha_{\pi\eta}^{\text{(B22)}}}{f^4},\frac{\alpha_{\eta\pi}^{\text{(B22)}}}{f^4}\Big\}\frac{g^2}{d-1}\Big\{(d^2-1)J_{41}^{B}+(J_{22}^{B}+2J_{32}^{B}+J_{43}^{B})\bm{q}^4\nonumber\\
&-[(\bm{\sigma}\cdot\bm{T}+1)J_{21}^{B}+2(d+1)(J_{31}^{B}+J_{42}^{B})]\bm{q}^2\Big\}\{(m_\pi,m_\pi,\delta_D),(m_K,m_K,\delta_D),\nonumber\\
&(m_\eta,m_\eta,\delta_D),(m_\pi,m_\eta,\delta_D),(m_\eta,m_\pi,\delta_D)\},
\end{align}
\begin{align}
\label{eq31}
\mathcal{V}_{BD^{*}}^{\text{(B23)}}=&\Big\{\frac{\alpha_{\pi\pi}^{\text{(B23)}}}{f^4},\frac{\alpha_{KK}^{\text{(B23)}}}{f^4},\frac{\alpha_{\eta\eta}^{\text{(B23)}}}{f^4},\frac{\alpha_{\pi\eta}^{\text{(B23)}}}{f^4},\frac{\alpha_{\eta\pi}^{\text{(B23)}}}{f^4}\Big\}g^2\mathcal{C}^2\frac{(d-3)(d-2)^2}{(d-1)^2}\Big\{(d^2-1)J_{41}^{B}+(J_{22}^{B}+\nonumber\\
&2J_{32}^{B}+J_{43}^{B})\bm{q}^4+\Big[\frac{\bm{\sigma}\cdot\bm{T}}{(d-2)^2}J_{21}^{B}-J_{21}^{B}-2(d+1)(J_{31}^{B}+J_{42}^{B})\Big]\bm{q}^2\Big\}\{(m_\pi,m_\pi,-\delta_B),\nonumber\\
&(m_K,m_K,-\delta_B),(m_\eta,m_\eta,-\delta_B),(m_\pi,m_\eta,-\delta_B),(m_\eta,m_\pi,-\delta_B)\},
\end{align}
\begin{align}
\label{eq32}
\mathcal{V}_{BD^{*}}^{\text{(B24)}}=&\Big\{\frac{\alpha_{\pi\pi}^{\text{(B24)}}}{f^4},\frac{\alpha_{KK}^{\text{(B24)}}}{f^4},\frac{\alpha_{\eta\eta}^{\text{(B24)}}}{f^4},\frac{\alpha_{\pi\eta}^{\text{(B24)}}}{f^4},\frac{\alpha_{\eta\pi}^{\text{(B24)}}}{f^4}\Big\}g^2\mathcal{C}^2\frac{d-2}{(d-1)^2}\Big\{(J_{22}^{B}+2J_{32}^{B}+J_{43}^{B})\bm{q}^4\nonumber\\
&+(d^2-1)J_{41}^{B}+\Big[\Big(\frac{\bm{\sigma}\cdot\bm{T}}{d-2}-1\Big)J_{21}^{B}-2(d+1)(J_{31}^{B}+J_{42}^{B})\Big]\bm{q}^2\Big\}\{(m_\pi,m_\pi,-\delta_B,\delta_D),\nonumber\\
&(m_K,m_K,-\delta_B,\delta_D),(m_\eta,m_\eta,-\delta_B,\delta_D),(m_\pi,m_\eta,-\delta_B,\delta_D),(m_\eta,m_\pi,-\delta_B,\delta_D)\},
\end{align}
\begin{align}
\label{eq33}
\mathcal{V}_{BD^{*}}^{\text{(R21)}}=\mathcal{V}_{BD^{*}}^{\text{(B21)}}|_{J_{ij}^{B}\rightarrow J_{ij}^{R},\alpha^{\text{(B21)}}_{**}\rightarrow\alpha^{\text{(R21)}}_{**},\bm{\sigma}\cdot\bm{T}\rightarrow-\bm{\sigma}\cdot\bm{T}},
\end{align}
\begin{align}
\label{eq34}
\mathcal{V}_{BD^{*}}^{\text{(R22)}}=\mathcal{V}_{BD^{*}}^{\text{(B22)}}|_{J_{ij}^{B}\rightarrow J_{ij}^{R},\alpha^{\text{(B22)}}_{**}\rightarrow\alpha^{\text{(R22)}}_{**},\bm{\sigma}\cdot\bm{T}\rightarrow-\bm{\sigma}\cdot\bm{T}},
\end{align}
\begin{align}
\label{eq35}
\mathcal{V}_{BD^{*}}^{\text{(R23)}}=\mathcal{V}_{BD^{*}}^{\text{(B23)}}|_{J_{ij}^{B}\rightarrow J_{ij}^{R},\alpha^{\text{(B23)}}_{**}\rightarrow\alpha^{\text{(R23)}}_{**},\bm{\sigma}\cdot\bm{T}\rightarrow-\bm{\sigma}\cdot\bm{T}},
\end{align}
\begin{align}
\label{eq36}
\mathcal{V}_{BD^{*}}^{\text{(R24)}}=\mathcal{V}_{BD^{*}}^{\text{(B24)}}|_{J_{ij}^{B}\rightarrow J_{ij}^{R},\alpha^{\text{(B24)}}_{**}\rightarrow\alpha^{\text{(R24)}}_{**},\bm{\sigma}\cdot\bm{T}\rightarrow-\bm{\sigma}\cdot\bm{T}},
\end{align}
where the parameters $\alpha^{(**)}_{**}$ are also detailed in the Appendix~\ref{AppParameters}. In the above expressions, all physical quantities are defined in $d$ dimensions, such as $\bm{S}^2=(d-1)/4$ and $q_iq_j=1/(d-1)\bm{q}^2\delta_{ij}$ for the $S$ wave.

At the next-to-leading order, in addition to the previously mentioned two-pseudoscalar-meson-exchange potentials, there exist one-loop corrections pertaining to the one-pseudoscalar-meson-exchange (refer to Fig.~\ref{fig:BDstarOneMesonOneloop}) and contact terms. These one-loop corrections can be incorporated by substituting the physical values of relevant parameters such as the couplings, decay constant, and masses of light-mesons, baryons, and heavy-mesons. Actually, the pseudoscalar-meson decay constant $f$ should be substituted by
\begin{align}
\label{fpifkfeta}
f_\pi=&f\Big[1-2\mu_\pi-\mu_K+\frac{4m_\pi^2}{f^2}L_5^r+\frac{8m_K^2+4m_\pi^2}{f^2}L_4^r\Big],\nonumber\\
f_K=&f\Big[1-\frac{3}{4}\mu_\pi-\frac{3}{2}\mu_K-\frac{3}{4}\mu_\eta+\frac{4m_K^2}{f^2}L_5^r+\frac{8m_K^2+4m_\pi^2}{f^2}L_4^r\Big],\nonumber\\
f_\eta=&f\Big[1-3\mu_K+\frac{4m_\eta^2}{f^2}L_5^r+\frac{8m_K^2+4m_\pi^2}{f^2}L_4^r\Big],
\end{align}
where
\begin{align}
\mu_\phi=\frac{m_\phi^2}{32\pi^2f^2}\text{ln}\frac{m_\phi^2}{\lambda^2}.
\end{align}
Details on the renormalization of the pseudoscalar-meson  decay constant $f$ to one-loop order are elaborated in ref.~\cite{Gasser:1984gg}. Note that we utilize the physical values $f_{\phi}(\phi=\pi,K,\eta)$ of the pseudoscalar-meson decay constants in the expression for the respective pseudoscalar mesons. The method of renormalization is straightforward in this context. The renormalization procedure for other constants, such as $g$, remains the same. For a more detailed understanding of the renormalization method, see ref.~\cite{Epelbaum:2002gb}. However, there might be distinctions between these methods regarding the renormalization of the pseudoscalar-meson decay constant, which are further elaborated upon in Appendix~\ref{AppDecayC}.

In addition, we also have the corrections from the subleading contact Lagrangians involving some unrenormalized $\mathcal{O}(\epsilon^2)$ LECs. The contact terms with two powers of momenta in the following form \cite{Machleidt:2011zz}:
\begin{align}
 \mathcal{V}_{BD^{*}}^{(\text{NLO})}=&C_{1}\bm{q}^2+C_2\bm{k}^2+(C_3\bm{q}^2+C_4 \bm{k}^2)\bm{\sigma}\cdot\bm{T}+C_5[i\bm{S}\cdot(\bm{q}\times\bm{k})]+C_6(\bm{\sigma}\cdot\bm{q})(\bm{T}\cdot\bm{q})\nonumber\\
 &+C_7(\bm{\sigma}\cdot\bm{k})(\bm{T}\cdot\bm{k}),
\end{align}
\begin{align}
 \mathcal{V}_{BD}^{(\text{NLO})}=&C_{1}\bm{q}^2+C_2\bm{k}^2+C_5[i\bm{S}\cdot(\bm{q}\times\bm{k})],
\end{align}
where $\bm{k}=(\bm{p}_{B}+\bm{p}_D)/2$ is the average momentum and $\bm{S}=(\bm{S}_B+\bm{S}_{D^{*}})/2$ is the total spin. The $C_i$ are the combinations of the LECs from the next-to-leading order contact Lagrangians. Nevertheless, their specific relations are not essential for our current discussion.

Utilizing the momentum-space potentials $\mathcal{V}(\bm{q})$ derived above, we proceed with the subsequent Fourier transformation to acquire the effective potential $V(r)$ in coordinate space,
\begin{align}
\label{eq37}
V(\bm{r})=\int\frac{d^3\bm{q}}{(2\pi)^3}e^{i\bm{q}\cdot\bm{r}}\mathcal{V}(\bm{q})\mathcal{F}(\bm{q}).
\end{align}
In this context, the introduction of a regulator function $\mathcal{F}(\bm{q})$ becomes essential to suppress the high momentum contributions. For this purpose, we opt for the Gaussian form $\mathcal{F}(\bm{q}) = \exp\left(-\bm{q}^{2n}/\mathit{\Lambda}^{2n}\right)$, previously utilized in studies related to the $NN$ and $N\bar{N}$ systems \cite{Epelbaum:2003xx,Epelbaum:2004fk,Kang:2013uia}. The Taylor expansion of this regulator function can be expressed as $\mathcal{F}(\bm{q}) = 1-\bm{q}^{2n}/\mathit{\Lambda}^{2n}+\cdots$. The choice of the power $n$ is crucial to ensure that the contributions from the cutoff-induced terms, $\mathcal{V}(\bm{q})\mathcal{O}(\bm{q}^{2n}/\mathit{\Lambda}^{2n})$, lie beyond the chiral order at which the analysis is conducted. Although our calculations primarily focus on the leading and subleading contributions, choosing $n=2$ would suffice. However, for consistency with the estimation of the LECs in the $N\bar{N}$ system \cite{Kang:2013uia}, where $n=3$ was adopted, we also select $n=3$. For the cutoff parameter $\mathit{\Lambda}$, we can make adjustments within the range of $400-600$ MeV, guided by insights from alternative analyses \cite{Entem:2003ft,Epelbaum:2014efa,Kang:2013uia,Dai:2017ont,Meng:2019nzy,Wang:2019nvm}. It is pretty obvious that the result for the potential may depend sensitively on the regulator and its cutoff parameter. As detailed in the introduction, the result is, in fact, independent of the regulator function, owing to the presence of contact terms. For further insights into the impact of the regulator function on the result, see Appendix~\ref{AppRegulator}.

Due to $\delta_D>m_\pi$, certain diagrams in Fig.~\ref{fig:BDstaroneloop}, specifically (T22) and (B22), would yield imaginary components, influencing the width of the associated bound state. If one were to solve the Lippmann-Schwinger equation, the presence of these imaginary parts would alter the pole position within the Riemann sheet. In this work, we opt for solving the Schr{\"o}dinger equation: 
\begin{align}
\label{eq38}
\Big[-\frac{\hbar^2}{2\mu}\nabla^2+V(\bm{r})\Big]\psi(\bm{r})=E\psi(\bm{r}),
\end{align}
where $\mu$ represents the reduced mass of system. Our study focuses primarily on binding energies. As a result, we exclude the imaginary components from our analysis, concentrating on the real dynamics of the system without the complexity added by these components.

\section{Results and discussion}
\label{results} 
To derive the numerical results, the $\mathcal{O}(\epsilon^0)$ LECs ($D_a, D_b, E_a, E_b)$ and the $\mathcal{O}(\epsilon^2)$ LECs $C_i$ need to be determined. Following the methodology proposed in ref.~\cite{Meng:2019nzy,Wang:2019nvm,Wang:2020dhf}, we estimate the LECs through the development of a contact Lagrangian at the quark level. The $\mathcal{O}(\epsilon^0)$ LECs involve the coupling constants $c_s$ and $c_t$. The relationships ($D_a=3c_s, D_b=-c_t, E_a=3c_s, E_b=-5c_t$) were deduced by aligning the $ND^{*}$ potentials from the hadron and quark levels in the SU(2) framework, as detailed in ref.~\cite{Wang:2020dhf}. Similarly, for the SU(3) framework, the relationships are established as ($D_a=2c_s, D_b=2c_t/3, E_a=3c_s, E_b=-5c_t$). The values of $c_s$ and $c_t$ can be estimated from the $N\bar{N}$ interaction, as detailed in ref.~\cite{Wang:2020dhf}. Undoubtedly, this method will introduce some uncertainties for the $\mathcal{O}(\epsilon^0)$ LECs. We use the values of $\tilde{C}_{^3 S_1}$ in the $I=0$ and $I=1$ channels at NLO and NNLO from ref.~\cite{Kang:2013uia}. Then we obtain $c_s=-8.2^{+1.3}_{-0.9}\pm 2.5\,\text{GeV}^{-2}$ and $c_t=1.1^{+1.2}_{-1.9}\pm 0.3\,\text{GeV}^{-2}$. The first uncertainty for the respective parameter is purely statistical. Due to the connection between these parameters and those of $N\bar{N}$ interaction, established using a simple quark model, we assign a 30\% systematic uncertainty to $c_s/c_t$ as its second uncertainty. One reason for the systematic uncertainty introduced by this matching process is that the quark-level parameters $c_s$ and $c_t$ are independent of $\mathit{\Lambda}$, while the LECs  $D_{a,b}$ and $E_{a,b}$ depend on $\mathit{\Lambda}$. Additionally, the LECs arising from the $N\bar{N}$ interaction also depend on $\mathit{\Lambda}$. Although reasonable results may be obtained for a particular value of $\mathit{\Lambda}$, the differing dependencies on $\mathit{\Lambda}$ can lead to a systematic uncertainty. Unfortunately, we do not have enough data to determine the values of $C_i$. However, based on the $\mathcal{O}(\epsilon^2)$ LECs of $N\bar{N}$ interaction, we estimate the $\mathcal{O}(\epsilon^2)$ LECs combined by $C_i$ ($C_{\text{NLO}}$) to be roughly between $-100 \,\text{GeV}^{-4}$ and $+100 \,\text{GeV}^{-4}$ for all channels, to the maximum extent possible.

Furthermore, we use the average values for the isospin multiplet masses of mesons and baryons from PDG \cite{ParticleDataGroup:2022pth}: $m_\pi=137.27\,\text{MeV}$, $m_K=495.64\,\text{MeV}$, $m_\eta=547.86\,\text{MeV}$, $M_D=1867.24\,\text{MeV}$, $M_{D_s}=1968.34\,\text{MeV}$, $M_{D^{*}}=2008.55\,\text{MeV}$, $M_{{D_s}^{*}}=2112.20\,\text{MeV}$, $M_N=938.92\,\text{MeV}$, $M_\Sigma=1193.15\,\text{MeV}$, $M_\Xi=1318.28\,\text{MeV}$, $M_\Lambda=1115.68\,\text{MeV}$, $M_\Delta=1232.00\,\text{MeV}$, $M_{\Sigma^{*}}=1384.57\,\text{MeV}$, $M_{\Xi^{*}}=1533.40\,\text{MeV}$, $M_\Omega=1672.45\,\text{MeV}$. We also consider the average values for the mass differences: $\delta_B=314.10\,\text{MeV}$ and $\delta_D=142.59\,\text{MeV}$. The physical values of the pseudoscalar decay constants, as reported in PDG, are used: $f_\pi=92.07\,\text{MeV}$, $f_K=110.03\,\text{MeV}$, and $f_\eta=119.69\,\text{MeV}$. For the results of selecting different decay constants, see Appendix~\ref{AppDecayC}. The axial vector coupling constant $g_A$ has been determined to be approximately 1.27, based on lattice quantum chromodynamics calculations \cite{Chang:2018uxx} and measurements in the decay of free neutrons \cite{Markisch:2018ndu}. Accordingly, we adopt $D=0.80$ and $F=0.47$ as their physical values. These parameters are crucial for obtaining accurate numerical results in our theoretical framework. 

In our analysis, we adopt $\mathcal{C}=1.2$ based on the study of strong and electromagnetic decays of decuplet baryons, as detailed in ref.~\cite{Butler:1992pn}. The coupling constant $g=-0.59$ is determined from the decay width of $D^{*+}$, as reported in refs.~\cite{CLEO:2001foe,Isola:2003fh}. The negative sign for $g$ is derived using insights from the quark model. Regarding the cutoff parameter $\mathit{\Lambda}$, an extensive discussion is presented in ref.~\cite{Wang:2020dhf}. The findings from this reference suggest that a cutoff value of $\mathit{\Lambda}=0.4\,\text{GeV}$ is suitable for accurately describing bound states. Consequently, we opt for $\mathit{\Lambda}=0.4\,\text{GeV}$. For the impact of the cutoff $\mathit{\Lambda}$ on the result, see Appendix~\ref{AppRegulator}.

\subsection{\texorpdfstring{$D_{(s)}^{(*)}N$}{} systems}

\begin{figure}[t]
\centering
\includegraphics[height=10.5cm,width=14.5cm]{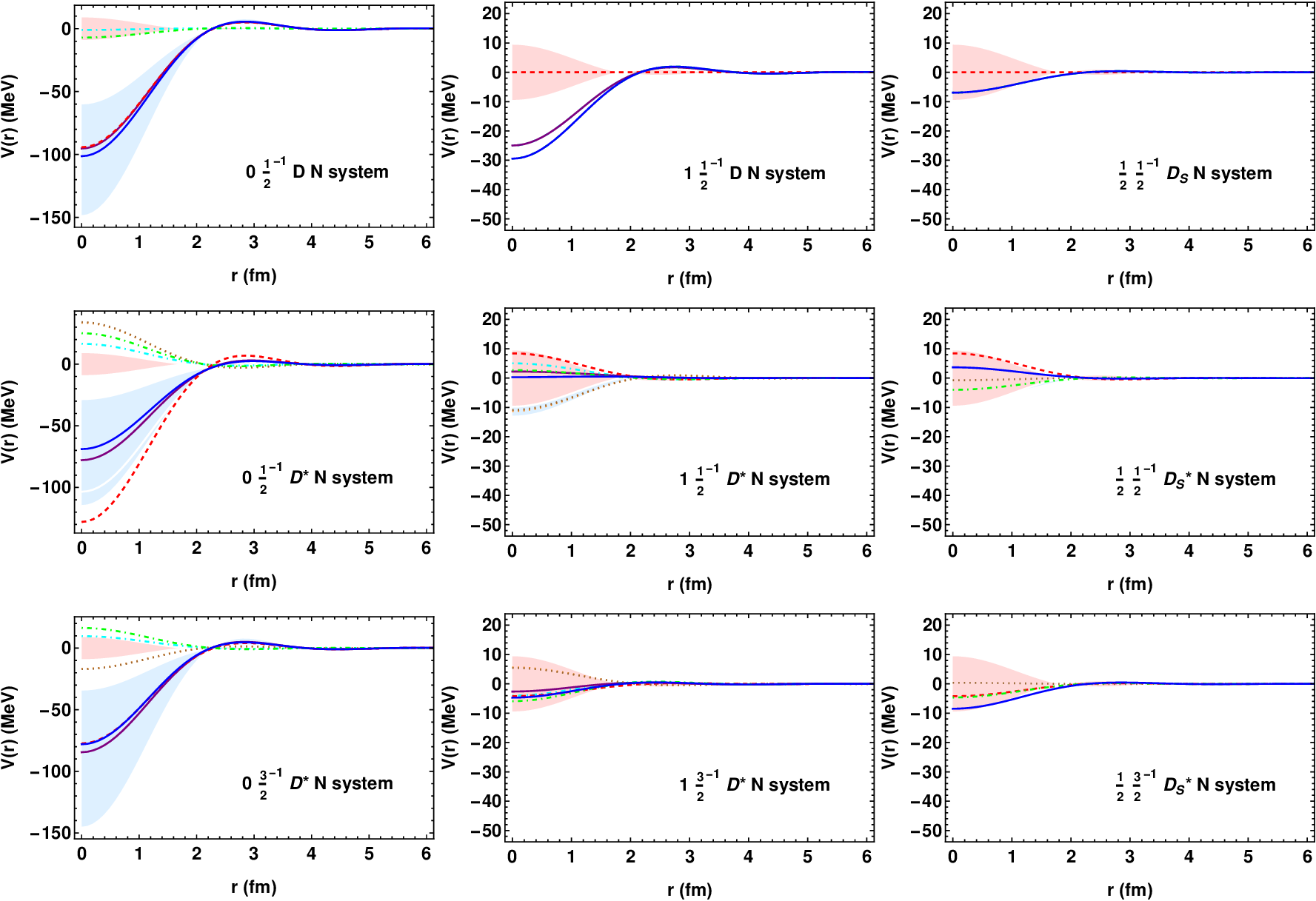}
\caption{\label{fig:DmesonNucleon}The effective potentials of the $D_{(s)}^{(*)}N$ systems. Their quantum numbers ($IJ^{P}$) are indicated in each subfigure. The red-dashed, brown-dotted, green-dotdashed, blue-solid lines denote the $\mathcal{O}(\epsilon^0)$ contact, one-meson-exchange, two-meson-exchange, and  their total potentials in SU(3) framework, respectively. Meanwhile, the pink-dashed, orange-dotted, cyan-dotdashed, purple-solid lines present the $\mathcal{O}(\epsilon^0)$ contact, one-meson-exchange, two-meson-exchange, and  their total potentials in SU(2) framework, respectively. The light-blue bands are estimated from the statistical uncertainties of the $\mathcal{O}(\epsilon^0)$ LECs using the standard error propagation formula, while the light-red bands denote the uncertainties of the $\mathcal{O}(\epsilon^2)$ LECs. }
\end{figure}

We first analyze the $D$-meson and nucleon systems. The effective potentials of each possible $IJ^P$ configuration are shown in Fig.~\ref{fig:DmesonNucleon}, where we present the $\mathcal{O}(\epsilon^0)$ contact, one-meson-exchange, two-meson-exchange, and their total potentials. Regarding the $\mathcal{O}(\epsilon^2)$ contact potentials, we provide them as error bands. We can observe that the maximum variation of these potentials is from -10 MeV to 10 MeV, which is relatively small and does not affect the bound states significantly. Therefore, it is reasonable to consider them as errors. For the $[DN]^{I=0}_{J=1/2}$ system, the $\mathcal{O}(\epsilon^0)$ contact and two-meson-exchange potentials are both attractive. However, the attraction of the two-meson-exchange potentials is relatively weak. The dominant source of the attractive potential arises from the $\mathcal{O}(\epsilon^0)$ contact interaction. For the $[DN]^{I=1}_{J=1/2}$ and $[D_s N]^{I=1/2}_{J=1/2}$ systems, the $\mathcal{O}(\epsilon^0)$ contact interactions vanish in our calculation, and the total potentials stem from the two-meson-exchange contributions. The potentials in the two channels exhibit considerably shallower depths when compared to the $[DN]^{I=0}_{J=1/2}$ channel, with a particular emphasis on the $[D_s N]^{I=1/2}_{J=1/2}$ channel.

In the $[D^{*}N]^{I=0}_{J=1/2}$ system, the $\mathcal{O}(\epsilon^0)$ contact interaction exhibits an attractive potential, while both the one-meson-exchange and two-meson-exchange potentials are repulsive. Despite this, the total potential of the system maintains approximately half the depth compared to that observed in the $[DN]^{I=0}_{J=1/2}$ channel. For the channel $[D^{*}N]^{I=0}_{J=3/2}$, the behavior of its potentials is intriguing. It is noteworthy that the contributions from one-meson-exchange and two-meson-exchange nearly offset each other. Consequently, the total potential is predominantly determined by the $\mathcal{O}(\epsilon^0)$ contact term, reaching a maximum depth of $-80$ MeV. In the $[D^{*}N]^{I=1}_{J=1/2}$, $[D^{*}N]^{I=1}_{J=3/2}$, $[D^{*}_{s}N]^{I=1/2}_{J=1/2}$, and $[D^{*}_{s}N]^{I=1/2}_{J=3/2}$ systems, the $\mathcal{O}(\epsilon^0)$ contact, one-meson-exchange, and two-meson-exchange potentials exhibit weakly attractive or weakly repulsive behavior. Consequently, their total potentials are also either weakly attractive or weakly repulsive. 

In the $[DN]^{I=0,1}_{J=1/2}$ and $[D^{*}N]^{I=0,1}_{J=1/2,3/2}$ channels, we present the potentials arising from the SU(2) framework, where the $\eta$- and $K$-exchange contributions are excluded. In both the SU(2) and SU(3) frameworks, the $\mathcal{O}(\epsilon^0)$ contact and one-meson potentials exhibit nearly identical characteristics, while a small distinction emerges in the two-meson potentials. The total potentials from the SU(3) and SU(2) frameworks are not significantly different. Therefore, our calculation in SU(3) framework is reasonable.

\begin{table*}[!t]
\centering
\begin{threeparttable}
\caption{\label{tab:DmesonNucleon}The bound states for the $D^{(*)}_{(s)}N$ systems at LO and NLO in SU(3) and SU(2) frameworks. The $\Delta E$, $M$, and $\sqrt{\left\langle r^2 \right\rangle}$ denote the binding energy, the mass of the bound state, and the root mean square radius, respectively. The first uncertainties are estimated from the $\mathcal{O}(\epsilon^0)$ LECs using the standard error propagation formula, while the second uncertainties are from the $\mathcal{O}(\epsilon^2)$ LECs.  }
\begin{tabular}{ccccccccccccccccccc}
\midrule \toprule
 \makecell[c]{SU(3) ~\\ \\ SU(2)} & $[DN]^{I=0}_{J=1/2}$ & $[D^{*}N]^{I=0}_{J=1/2}$ & $[D^{*}N]^{I=0}_{J=3/2}$& \\
\midrule
$\text{LO}:\,\Delta E$ ($\text{MeV}$) & \makecell[c]{$-10.8^{+7.4}_{-7.2}$ ~\\ \\ $-10.8^{+7.4}_{-7.2}$} &\makecell[c]{$-12.3^{+11.8}_{-5.9}$~\\ \\ $-12.5^{+11.8}_{-5.9}$} &\makecell[c]{$-10.9^{+10.9}_{-20.2}$~\\ \\ $-10.9^{+10.9}_{-20.2}$} &\\
\midrule
$\text{LO}:\,M$ ($\text{MeV}$) &\makecell[c]{$2795.4^{+7.4}_{-7.2}$ ~\\ \\ $2795.4^{+7.4}_{-7.2}$} &\makecell[c]{$2935.1^{+11.8}_{-5.9}$ ~\\ \\ $2935.0^{+11.8}_{-5.9}$} &\makecell[c]{$2936.5^{+10.9}_{-20.2}$~\\ \\ $2936.5^{+10.9}_{-20.2}$}  &\\
\midrule
$\text{LO}:\,\sqrt{\left\langle r^2 \right\rangle}$ ($\text{fm}$) &\makecell[c] {$1.8^{+0.7}_{-0.3}$ ~\\ \\ $1.8^{+0.7}_{-0.3}$}  &\makecell[c]{$1.8^{+1.4}_{-0.3}$~\\ \\ $1.8^{+1.3}_{-0.3}$} & \makecell[c]{$1.8^{+2.5}_{-0.5}$~\\ \\ $1.8^{+2.5}_{-0.5}$} &\\
\midrule
$\text{NLO}:\,\Delta E$ ($\text{MeV}$) &\makecell[c]{$-13.4^{+8.8}_{-8.1}\pm 3.2$~\\ \\ $-10.9^{+8.0}_{-7.8}\pm 3.1 $}& \makecell[c]{$-4.5^{+4.5}_{-4.6}\pm 2.1$~\\ \\$-7.1^{+7.1}_{-5.2}\pm 2.4$} &\makecell[c]{$-5.0^{+5.0}_{-19.1}\pm 2.5$ ~\\ \\ $-7.2^{+7.2}_{-20.2}\pm 2.8$} &\\
\midrule
$\text{NLO}:\,M$ ($\text{MeV}$) &\makecell[c]{$2792.8^{+8.8}_{-8.1}\pm 3.2$ ~\\ \\ $2795.1^{+8.0}_{-7.8}\pm 3.1$} &\makecell[c]{$2942.9^{+4.5}_{-4.6}\pm 2.1$~\\ \\ $2940.4^{+7.1}_{-5.2}\pm 2.4$} &\makecell[c]{$2942.4^{+5.0}_{-19.1}\pm 2.5$ ~\\ \\ $2940.2^{+7.2}_{-20.2}\pm 2.8$}  &\\
\midrule
$\text{NLO}:\,\sqrt{\left\langle r^2 \right\rangle}$ ($\text{fm}$) &\makecell[c]{$1.7^{+0.6}_{-0.3}\pm 0.1$~\\ \\ $1.8^{+0.7}_{-0.3}\pm 0.1$}  & \makecell[c]{$2.5^{+9.4}_{-0.6}\pm 0.3$ ~\\ \\ $2.1^{+4.9}_{-0.3}\pm 0.2$} &\makecell[c]{$2.3^{+12.7}_{-1.1}\pm 0.3$ ~\\ \\ $2.0^{+8.5}_{-0.7}\pm 0.2$} &\\
\bottomrule \midrule
\end{tabular} 
\end{threeparttable}
\end{table*}

We obtain the binding solutions in the $[DN]^{I=0}_{J=1/2}$, $[D^{*}N]^{I=0}_{J=1/2}$, and $[D^{*}N]^{I=0}_{J=3/2}$ systems at LO and NLO, and the results are detailed in Table~\ref{tab:DmesonNucleon}. The first uncertainties are estimated from the $\mathcal{O}(\epsilon^0)$ LECs using the standard error propagation formula, while the second uncertainties for the NLO results are from the $\mathcal{O}(\epsilon^2)$ LECs. These second uncertainties are relatively small, suggesting that it is reasonable to treat the $\mathcal{O}(\epsilon^2)$ contact contributions as errors. It also indicates that the results for bound states are not very sensitive to the $\mathcal{O}(\epsilon^2)$ contact terms. The binding energy values for the SU(2) framework exhibit slightly deviations from those reported in ref.~\cite{Wang:2020dhf}. This discrepancy primarily arises from our selection of different constants, particularly a smaller coupling constants for the $\Delta N \pi$ vertex. A key point of interest is the $[D^{*}N]^{I=0}_{J=1/2,3/2}$ channels. The masses of the two bound states closely align with the $\Lambda_c(2940)$ mass measurements. Additionally, the results for the bound states at LO and NLO are consistent within errors. This indicates that our calculations have very good convergence. It is important to note that the three bound states have significant uncertainties, and their existence is not definitively established. More precise calculations are required.

\subsection{\texorpdfstring{$D_{(s)}^{(*)}\Sigma$}{} systems}

\begin{figure}[t]
\centering
\includegraphics[height=10.5cm,width=14.5cm]{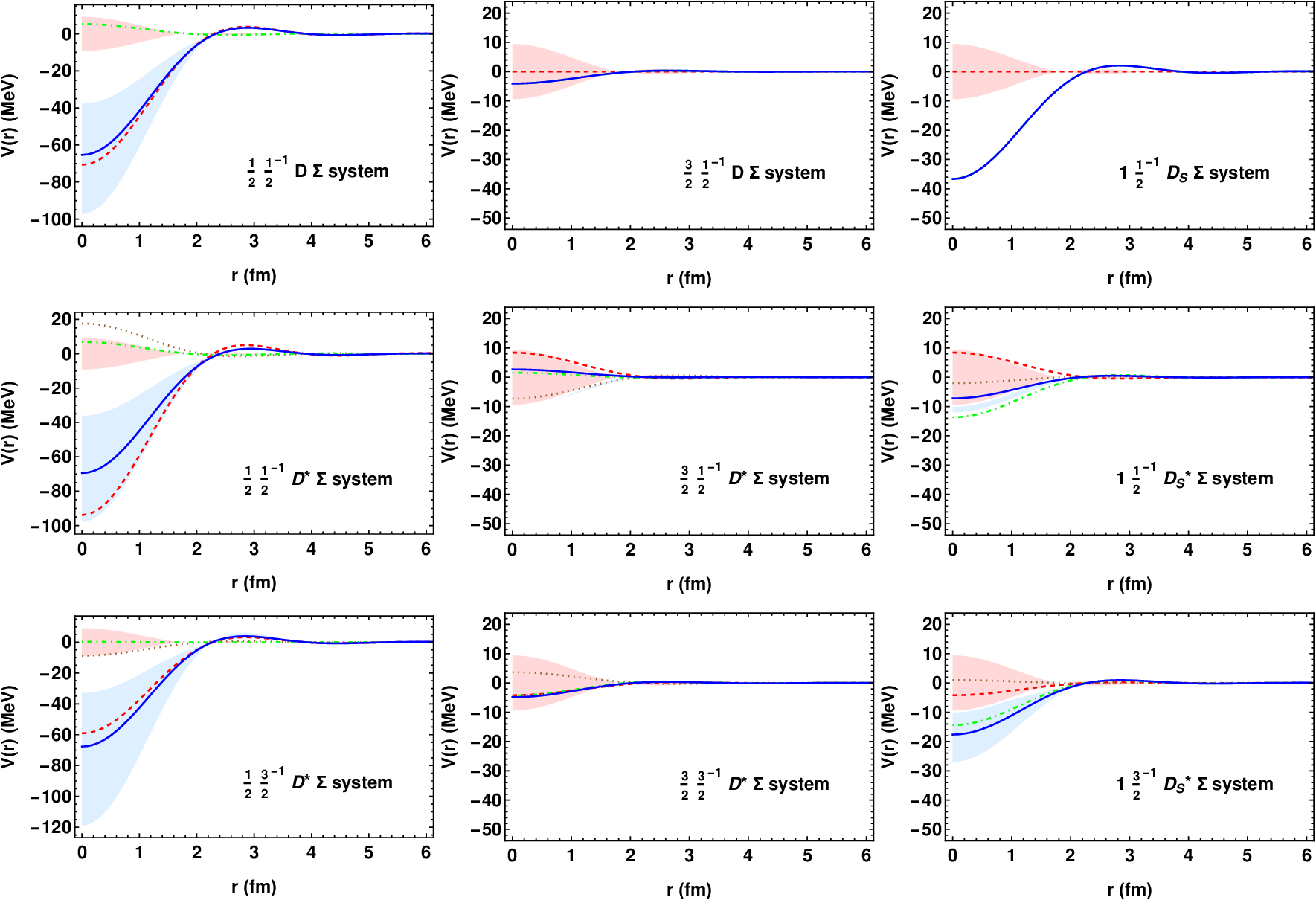}
\caption{\label{fig:DmesonSigma}The effective potentials of the $D_{(s)}^{(*)}\Sigma$ systems. The notation is the same as in Fig.~\ref{fig:DmesonNucleon}.}
\end{figure}

\begin{table*}[b]
\centering
\begin{threeparttable}
\caption{\label{tab:DmesonSigma}The bound states for the $D_{(s)}^{(*)}\Sigma$ systems at LO and NLO in SU(3) framework. The notation is the same as in Table~\ref{tab:DmesonNucleon}.}
\begin{tabular}{ccccccccccccccccccc}
\midrule \toprule
 SU(3) & $[D\Sigma]^{I=1/2}_{J=1/2}$ & $[D^{*}\Sigma]^{I=1/2}_{J=1/2}$ & $[D^{*}\Sigma]^{I=1/2}_{J=3/2}$& \\
 \midrule
\text{LO}:\,$\Delta E$ ($\text{MeV}$)&$-5.6^{+5.4}_{-5.4}$&$-8.7^{+8.7}_{-3.8}$ &$-4.9^{+4.9}_{-14.3}$ &\\
\midrule
\text{LO}:\,$M$ ($\text{MeV}$)&$3054.8^{+5.4}_{-5.4}$ &$3193.0^{+8.7}_{-3.8}$ &$3196.8^{+4.9}_{-14.3}$  &\\
\midrule
\text{LO}:\,$\sqrt{\left\langle r^2 \right\rangle}$ ($\text{fm}$) & $2.1^{+1.5}_{-0.5}$  & $1.9^{+1.4}_{-0.3}$ & $2.2^{+8.4}_{-0.9}$ &\\
\midrule
\text{NLO}:\,$\Delta E$ ($\text{MeV}$)&$-4.3^{+4.3}_{-4.9}\pm 2.3$&$-6.8^{+6.8}_{-3.5}\pm 2.7$ &$-4.9^{+4.9}_{-14.1} \pm 2.5 $  &\\
\midrule
\text{NLO}:\,$M$ ($\text{MeV}$)&$3056.1^{+4.3}_{-4.9}\pm 2.3$ &$3194.9^{+6.8}_{-3.5}\pm 2.7$ &$3196.8^{+4.9}_{-14.1} \pm 2.5 $  &\\
\midrule
\text{NLO}:\,$\sqrt{\left\langle r^2 \right\rangle}$ ($\text{fm}$) & $2.4^{+2.1}_{-0.8}\pm 0.4$  & $2.0^{+2.1}_{-0.2}\pm 0.2$ & $2.2^{+8.3}_{-0.9} \pm 0.3 $&\\
\bottomrule \midrule
\end{tabular}
\end{threeparttable}
\end{table*}

We then examine the $D$-meson and $\Sigma$-baryon systems. The effective potentials corresponding to each possible $IJ^P$ configuration are depicted in Fig.~\ref{fig:DmesonSigma}. The features of $D\Sigma$ are very similar to those of $DN$, with the distinction that the effective potential of $[D_s \Sigma]_{J=1/2}^{I=1}$ is deeper than that of $[D\Sigma]_{J=1/2}^{I=3/2}$, while the effective potential of $[D_s N]_{J=1/2}^{I=1/2}$ is shallower than that of $[DN]_{J=1/2}^{I=1}$. This discrepancy is anticipated, given the presence of an attractive $s\bar{s}$ quark pair in $[D_s \Sigma]_{J=1/2}^{I=1}$. Despite the effective potential of $[D\Sigma]^{I=1/2}_{J=1/2}$ being shallower than that of $[DN]^{I=0}_{J=1/2}$, it remains sufficiently deep to facilitate the formation of a bound state.

The characteristics of the $D^{*}\Sigma$ system show a striking similarity to those of the $D^{*}N$ system. For the low-isospin channels, $[D^{*}\Sigma]^{I=1/2}_{J=1/2,3/2}$, we observe significantly deep attractive potentials. Conversely, the high-isospin channels, $[D^{*}\Sigma]^{I=3/2}_{J=1/2,3/2}$, as well as the strange channels $[D_{s}^{*}\Sigma]^{I=1}_{J=1/2,3/2}$, are characterized by shallower attractive potentials. In the case of the $[D^{*}\Sigma]^{I=1/2}_{J=1/2}$ system, there is an attractive contact interaction, although both one-meson-exchange and two-meson-exchange potentials are repulsive. This particular trait mirrors that of the $[D^{*}N]^{I=0}_{J=1/2}$ system. In contrast, in the $[D^{*}\Sigma]^{I=1/2}_{J=3/2}$ system, the influence of the two-meson-exchange is negligible, diverging from the characteristics of the $[D^{*}N]^{I=0}_{J=3/2}$ system. Nevertheless, the attractive nature of the low-isospin channels is sufficient to facilitate the formation of bound states.

\begin{figure}[!t]
\centering
\includegraphics[height=10.5cm,width=14.5cm]{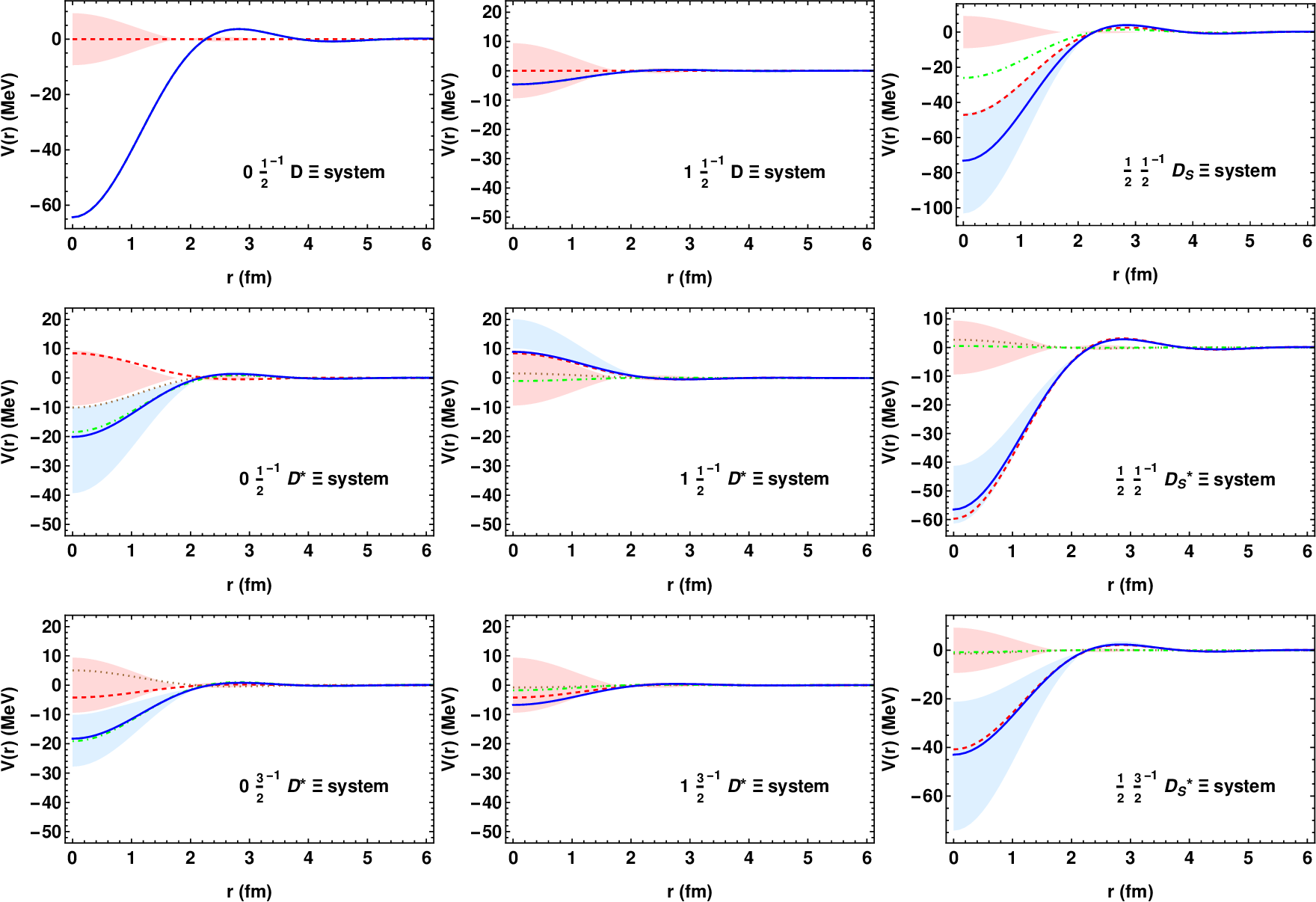}
\caption{\label{fig:DmesonXi}The effective potentials of the $D_{(s)}^{(*)}\Xi$ systems. The notation is the same as in Fig.~\ref{fig:DmesonNucleon}.}
\end{figure}

\begin{table*}[b]
\centering
\resizebox{\textwidth}{!}{
\begin{threeparttable}
\caption{\label{tab:DmesonXi}The bound states for the $D_{(s)}^{(*)}\Xi$ systems at LO and NLO in SU(3) framework. The notation is the same as in Table~\ref{tab:DmesonNucleon}.}
\begin{tabular}{ccccccccccccccccccc}
\midrule \toprule
 SU(3) & $[D\Xi]^{I=0}_{J=1/2}$ & $[D_{s}\Xi]^{I=1/2}_{J=1/2}$ & $[D_{s}^{*}\Xi]^{I=1/2}_{J=1/2}$&  $[D_{s}^{*}\Xi]^{I=1/2}_{J=3/2}$ &\\
\midrule
$\text{LO}:\,\Delta E$ ($\text{MeV}$)&$-$&$-0.6^{+0.6}_{-1.7}$ &$-3.2^{+2.7}_{-0.8}$ & $-0.1^{+0.1}_{-3.5}$ &\\
\midrule
$\text{LO}:\,M$ ($\text{MeV}$) &$-$ &$3286.0^{+0.6}_{-1.7}$  & $3427.2^{+2.7}_{-0.8}$ & $3430.4^{+0.1}_{-3.5}$  &\\
\midrule
$\text{LO}:\,\sqrt{\left\langle r^2 \right\rangle}$ ($\text{fm}$) & $-$  & $4.9^{+5.3}_{-2.3}$ & $2.5^{+1.7}_{-0.2}$ &  $7.5^{+5.3}_{-6.4}$ &\\
\midrule
$\text{NLO}:\,\Delta E$ ($\text{MeV}$)&$-4.1^{+0.0}_{-0.0}\pm 2.5$&$-7.9^{+3.8}_{-3.4}\pm 3.0$ &$-3.2^{+2.9}_{-0.7}\pm 2.1$ & $-0.2^{+0.2}_{-3.6}\pm 0.9$ &\\
\midrule
$\text{NLO}:\,M$ ($\text{MeV}$) &$3181.4^{+0.0}_{-0.0}\pm 2.5$ &$3278.7^{+3.8}_{-3.4}\pm 3.0$  & $3427.3^{+2.9}_{-0.7}\pm 2.1$ & $3430.3^{+0.2}_{-3.6}\pm 0.9$  &\\
\midrule
$\text{NLO}:\,\sqrt{\left\langle r^2 \right\rangle}$ ($\text{fm}$) & $2.3^{+0.0}_{-0.0}\pm 0.4$  & $1.9^{+0.4}_{-0.3}\pm 0.2 $ & $2.5^{+1.8}_{-0.2}\pm 0.4$ &  $7.1^{+5.6}_{-5.9}\pm 4.4$ &\\
\bottomrule \midrule
\end{tabular}
\end{threeparttable}}
\end{table*}

We also obtain binding solutions in the $[D\Sigma]^{I=1/2}_{J=1/2}$, $[D^{*}\Sigma]^{I=1/2}_{J=1/2}$, and $[D^{*}\Sigma]^{I=1/2}_{J=3/2}$ systems at LO and NLO. The numerical results with uncertainties are presented in Table~\ref{tab:DmesonSigma}. It is evident that the bound states from $D^{*}\Sigma$ exhibit similarities to those from $D^{*}N$. The channel with low spin, $[D^{*}\Sigma]^{I=1/2}_{J=1/2}$, has a almost same binding energy compared to the channel with high spin, $[D^{*}\Sigma]^{I=1/2}_{J=3/2}$. In analogy to interpreting $\Lambda_{c}(2940)$ as a molecular state of $D^{*}N$, we may propose to  interpret the $\Xi_{c}(3196)$ as a bound state of $D^{*}\Sigma$ with $IJ^P=\frac{1}{2}\frac{3}{2}^{-}$ or $\frac{1}{2}\frac{1}{2}^{-}$. It is noteworthy that the molecular states also emerge in the $[D\Sigma]^{I=1/2}_{J=1/2}$ and $[D^{*}\Sigma]^{I=1/2}_{J=3/2}$ systems from the calculations in ref.~\cite{Wang:2023eng}. Additionally, the $\Xi_{c}(3055)$ may be interpreted as the bound state of $[D\Sigma]^{I=1/2}_{J=1/2}$ system considering their nearly identical mass. Similar to the case of the $D_{(s)}^{(*)} N$ systems, the three bound states also have significant uncertainties, and their existence is not definitively established. Further precise calculations are needed.

\subsection{\texorpdfstring{$D_{(s)}^{(*)}\Xi$}{} systems}

We also investigate the systems composed of a $D$ meson and a $\Xi$ baryon. The effective potentials for each possible $IJ^P$ configuration are illustrated in Fig.~\ref{fig:DmesonXi}. A notable distinction emerges when comparing the characteristics of the $D\Xi$ systems to those of $DN$. Specifically, for the $[D\Xi]^{I=0}_{J=1/2}$ and $[D\Xi]^{I=1}_{J=1/2}$ systems, the $\mathcal{O}(\epsilon^0)$ contact interactions vanish in our calculation, since the total potentials are derived solely from the two-meson-exchange contributions. Conversely, in the $[D_{s}\Xi]^{I=1/2}_{J=1/2}$ system, the $\mathcal{O}(\epsilon^0)$ contact interaction is significant. This distinction leads to the emergence of sufficiently attractive potentials in the $[D\Xi]^{I=0}_{J=1/2}$ and $[D_{s}\Xi]^{I=1/2}_{J=1/2}$ systems, potentially indicating the formation of bound states. Meanwhile, a modestly attractive potential is observed in the $[D\Xi]^{I=1}_{J=1/2}$ system.

\begin{figure}[t]
\centering
\includegraphics[height=6.5cm,width=14.5cm]{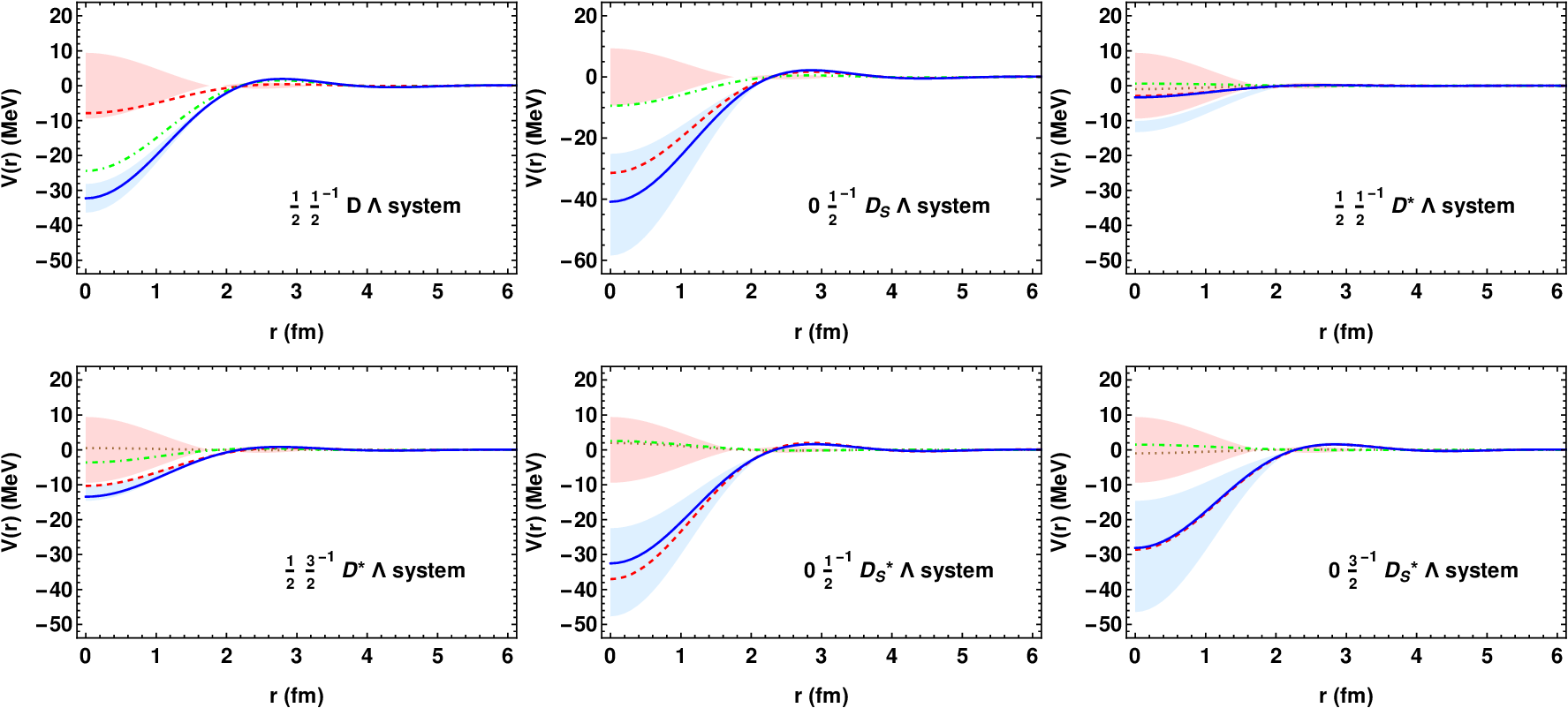}
\caption{\label{fig:DmesonLambda}The effective potentials of the $D_{(s)}^{(*)}\Lambda$ systems. The notation is the same as in Fig.~\ref{fig:DmesonNucleon}.}
\end{figure}

The $D^{*}\Xi$ system exhibits distinct characteristics compared to the $D^{*}N$ system. In the low-isospin channels, specifically $[D^{*}\Xi]^{I=0}_{J=1/2,3/2}$, we detect attractive potentials. However, these potentials are not sufficiently deep to facilitate the formation of bound states. Conversely, in the high-isospin channels, $[D^{*}\Xi]^{I=1}_{J=1/2,3/2}$, we observe a mixture of shallower repulsive and attractive potentials. With regard to the $[D_{s}^{*}\Xi]^{I=1/2}_{J=1/2,3/2}$ systems, both the one-meson-exchange and two-meson-exchange potentials appear relatively weak. The primary source of the attractive potential in these cases stem from the $\mathcal{O}(\epsilon^0)$ contact interactions. Despite the total potentials are not profoundly deep, they are sufficiently strong to potentially lead to the formation of bound states.

We obtain binding solutions in the $[D\Xi]^{I=0}_{J=1/2}$, $[D_{s}\Xi]^{I=1/2}_{J=1/2}$, $[D^{*}_{s}\Xi]^{I=1/2}_{J=1/2}$, and $[D^{*}_{s}\Xi]^{I=1/2}_{J=3/2}$ systems at LO and NLO, except for the $[D\Xi]^{I=0}_{J=1/2}$ system at LO due to the vanishing of its $\mathcal{O}(\epsilon^0)$ contact LECs,  with comprehensive results detailed in Table~\ref{tab:DmesonXi}. The bound state associated with the $[D^{*}_{s}\Xi]^{I=1/2}_{J=1/2,3/2}$ system may correspond to the $\Xi_{c}(3429)$.
Furthermore, the $\Xi_{c}(3279)$ may be interpreted as the bound state of the $[D_{s}\Xi]^{I=1/2}_{J=1/2}$. Intriguingly, its mass closely aligns with that of the bound state in the $[D^{*}\Sigma]^{I=1/2}_{J=1/2}$ system, and they share identical quantum numbers. These two channels may be considered as the coupled-channels. Additionally, the bound state in the $[D\Xi]^{I=0}_{J=1/2}$ system may correspond to the $\Omega_{c}(3188)$ observed by LHCb ~\cite{LHCb:2017uwr}. Our result for this channel is in agreement with ref.~\cite{Wang:2017smo}, where a bound state was identified. However, it contrasts with the findings in refs.~\cite{Chen:2017gnu,Wang:2023eng}, in which a $2S$ $\Omega_c$-baryon and a virtual state were reported, respectively. Unfortunately, the four bound states still have considerable uncertainties, and the reliability of this result will need to be confirmed by more precise calculations in the future.

\subsection{\texorpdfstring{$D_{(s)}^{(*)}\Lambda$}{} systems}

Finally, we explore the $D$-meson and $\Lambda$-baryon systems. For each possible $IJ^P$ configuration within these systems, we illustrate the effective potentials in Fig.~\ref{fig:DmesonLambda}. Compared with the previously discussed systems, it is noteworthy that all the $\mathcal{O}(\epsilon^0)$ contact interactions are nonvanishing here. However, these interactions are not sufficiently attractive to form bound states, even when taking into account both the one-meson-exchange and two-meson-exchange contributions. Particularly in the $[D_{s}\Lambda]^{I=0}_{J=1/2}$ system, we observe a significantly attractive contact potential coupled with an attractive two-meson-exchange potential, results in the most significant potential observed within the $D_{(s)}^{(*)}\Lambda$ systems. It is crucial to emphasize that no binding solution exists in these systems.

\section{Summary}
In this work, we have comprehensively calculated the effective potentials of the octet baryon and $D$-meson systems up to the next-to-leading order using the chiral effective field theory. Our approach encompasses the long-range one-pseudoscalar-meson-exchange potential, the mid-range two-pseudoscalar-meson-exchange potential, and the short-range contact interactions. Additionally, we have incorporated the decuplet baryons as the intermediate states within the loop diagrams. The involved low energy constants are judiciously estimated based on the $N\bar{N}$ interaction with help of the quark model.

In our investigation of the $D^{(*)}_{(s)}N$ systems, we have almost obtained attractive potentials in all channels, with the exception of the $[D_{s}^{*}N]^{I=1/2}_{J=1/2}$ channel. Significantly, bound states emerge only in the $[DN]^{I=0}_{J=1/2}$, $[D^{*}N]^{I=0}_{J=1/2}$, and $[D^{*}N]^{I=0}_{J=3/2}$ systems, owing to the sufficient depth of their attractive potentials. The masses of the bound states in the $[D^{*}N]^{I=0}_{J=1/2,3/2}$ systems exhibit a remarkable correspondence with the mass of the $\Lambda_{c}(2940)$. 

In our examination of the $D^{(*)}_{(s)}\Sigma$ systems, we observe that their characteristics closely mirror those of the $D^{(*)}_{(s)}N$ systems. We have obtained attractive potentials in all channels. Specifically, bound states appear in the $[D\Sigma]^{I=1/2}_{J=1/2}$, $[D^{*}\Sigma]^{I=1/2}_{J=1/2}$, and $[D^{*}\Sigma]^{I=1/2}_{J=3/2}$ systems. The bound state in the $[D\Sigma]^{I=1/2}_{J=1/2}$ system aligns in mass with $\Xi_{c}(3055)$, suggesting its interpretation as a molecular state. The bound state corresponding to $D^{*}\Sigma$ leads to the molecular explanation of the $\Xi_{c}(3196)$. These two cases parallel the interpretation of $\Lambda_{c}(2940)$ as a molecular state of $D^{*}N$. 

In our analysis of the $D^{(*)}_{(s)}\Xi$ systems, we observe distinct characteristics compared to the $D^{(*)}_{(s)}N$ systems. While a repulsive potential is identified in the $[D^{*}\Xi]^{I=1}_{J=1/2}$ channel, attractive potentials are present in the other channels. Significantly deep potentials, capable of generating bound states, are found in the $[D\Xi]^{I=0}_{J=1/2}$, $[D_{s}\Xi]^{I=1/2}_{J=1/2}$, $[D_{s}^{*}\Xi]^{I=1/2}_{J=1/2}$, $[D_{s}^{*}\Xi]^{I=1/2}_{J=3/2}$ systems. The bound state in the $[D\Xi]^{I=0}_{J=1/2}$ system may correspond to the $\Omega_{c}(3188)$. Moreover, the $\Xi_{c}(3279)$ and $\Xi_{c}(3429)$ may be interpreted as the molecular states of $[D_{s}\Xi]^{I=1/2}_{J=1/2}$ and $[D_{s}^{*}\Xi]^{I=1/2}_{J=1/2,3/2}$, respectively.

In our study of the $D^{(*)}_{(s)}\Lambda$ systems, attractive potentials are identified in all channels. Nevertheless, these attractive potentials are not sufficiently deep to facilitate the formation of bound states in any of these systems.

To summarize, our investigations provide new insights into the $\Lambda_{c}(2940)$, $\Xi_{c}(3055)$, and $\Omega_{c}(3188)$ states from the hadronic molecular perspective. We have also observed specific molecular states, referred to as $\Xi_{c}$, within the mass range of $3100-3500$ MeV. However, our calculations still have significant uncertainties, and the existence of these bound states will require more accurate calculations in the future, particularly when analyzed in conjunction with experimental results or lattice QCD. We look forward to future experimental validation of the possible existence of the singly heavy molecular baryon states. Such findings could potentially improve our understanding of the molecular state dynamics in heavy quark systems.

\section*{Acknowledgments}
This work is supported by the National Natural Science Foundation of China under Grants No. 11975033, No. 12070131001, No. 12147127, No. 12105072, and No. 12305090. B.-L. Huang is also supported by the Inner Mongolia Autonomous Region Natural Science Fund (No. 2024MS01016), the Research Support for High-Level Talent at the Autonomous Region Level (No. 13100-15112042) and the Junma Program High-Level Talent Reseach Start-up Fund (No. 10000-23112101/085). B. Wang was also supported by the Youth Funds of Hebei Province (No. A2021201027) and the Start-up Funds for Young Talents of Hebei University (No. 521100221021). We thank Zi-Yang Lin (Peking University), Lu Meng (Ruhr-Universit\"{a}t Bochum) for very helpful discussions.

\appendix\markboth{Appendix}{Appendix}
\renewcommand{\thesection}{\Alph{section}}
\numberwithin{equation}{section} \numberwithin{table}{section}

\section{Parameters for the potentials}
\label{AppParameters}
This appendix details the parameters for the potentials, as shown in Tables~\ref{constantsOne} to \ref{constantsTwo} below.

\begin{sidewaystable}
\centering
\resizebox{\textwidth}{!}{%
\begin{threeparttable}
\caption{\label{constantsOne}Parameters for the potentials of the $BD$ or $BD^*$ systems.}
\begin{tabular}{lcccccccccccc}
\midrule \toprule
Systems &$ND^{(*)}(I=1)$  & $ND^{(*)}(I=0)$ & $ND^{(*)}_{S}(I=\frac{1}{2})$ &$\Sigma D^{(*)}(I=\frac{3}{2})$  &$\Sigma D^{(*)}(I=\frac{1}{2})$ & $\Sigma D^{(*)}_{S}(I=1)$&$\Xi D^{(*)}(I=1)$ & $\Xi D^{(*)}(I=0)$ & $\Xi D^{(*)}_{S}(I=\frac{1}{2})$ & $\Lambda D^{(*)}(I=\frac{1}{2})$ & $\Lambda D^{(*)}_{S}(I=0)$ \\
\midrule
$\alpha^{\text{(X11)}}/\alpha^{\text{(X21)}}$&$-\frac{2}{3}$&$\frac{10}{3}$&$-\frac{2}{3}$&$-\frac{2}{3}$&$\frac{7}{3}$&$-\frac{2}{3}$&$-\frac{2}{3}$&$-\frac{2}{3}$&$\frac{4}{3}$&$-\frac{1}{3}$&$\frac{2}{3}$\\
\midrule
$\alpha^{\text{(H21)}}_{\pi}$&$\frac{D+F}{4}$&$-\frac{3}{4} (D+F)$&$0$&$\frac{F}{2}$&$-F$&$0$&$\frac{F-D}{4}$&$\frac{3 (D-F)}{4}$&$0$&$0$&$0$\\
\midrule
$\alpha^{\text{(H21)}}_{\eta}$&$\frac{1}{12} (D-3 F)$&$\frac{1}{12} (D-3 F)$&$\frac{1}{6} (3 F-D)$&$-\frac{D}{6}$&$-\frac{D}{6}$&$\frac{D}{3}$&$\frac{1}{12} (D+3 F)$&$\frac{1}{12} (D+3 F)$&$\frac{1}{6} (-D-3 F)$&$\frac{D}{6}$&$-\frac{D}{3}$\\
\midrule
$\alpha^{\text{(F11)}}_{\pi\pi}/\alpha^{\text{(F21)}}_{\pi\pi}$&$\frac{1}{4}$&$-\frac{3}{4}$&$0$&$\frac{1}{2}$&$-1$&$0$&$\frac{1}{4}$&$-\frac{3}{4}$&$0$&$0$&$0$\\
\midrule
$\alpha^{\text{(F11)}}_{KK}/\alpha^{\text{(F21)}}_{KK}$&$-\frac{3}{16}$&$-\frac{9}{16}$&$\frac{11}{16}$&$\frac{3}{16}$&$-\frac{3}{8}$&$\frac{1}{16}$&$\frac{3}{8}$&$0$&$-\frac{5}{8}$&$0$&$0$\\
\midrule
$\alpha^{\text{(T11)}}_{\pi\pi}$&$\frac{1}{4}$&$-\frac{3}{4}$&$0$&$\frac{1}{2}$&$-1$&$0$&$\frac{1}{4}$&$-\frac{3}{4}$&$0$&$0$&$0$\\
\midrule
$\alpha^{\text{(T11)}}_{KK}$&$-\frac{1}{4}$&$-\frac{3}{4}$&$\frac{3}{4}$&$\frac{1}{4}$&$-\frac{1}{2}$&$0$&$\frac{1}{2}$&$0$&$-\frac{3}{4}$&$0$&$0$\\
\midrule
$\alpha^{\text{(T12)}}_{\pi\pi}/\alpha^{\text{(T24)}}_{\pi\pi}$&$\frac{1}{6}$&$-\frac{1}{2}$&$0$&$-\frac{1}{24}$&$\frac{1}{12}$&$0$&$-\frac{1}{12}$&$\frac{1}{4}$&$0$&$0$&$0$\\
\midrule
$\alpha^{\text{(T12)}}_{KK}/\alpha^{\text{(T24)}}_{KK}$&$\frac{1}{12}$&$0$&$-\frac{1}{8}$&$-\frac{1}{12}$&$-\frac{5}{24}$&$\frac{1}{4}$&$-\frac{1}{24}$&$\frac{3}{8}$&$-\frac{1}{8}$&$\frac{1}{8}$&$-\frac{1}{4}$\\
\midrule
$\alpha^{\text{(T13)}}_{\pi\pi}/\alpha^{\text{(T23)}}_{\pi\pi}$&$\frac{1}{4} (D+F)^2$&$-\frac{3}{4} (D+F)^2$&$0$&$\frac{1}{6} \left(D^2+3 F^2\right)$&$-\frac{D^2}{3}-F^2$&$0$&$\frac{1}{4} (D-F)^2$&$-\frac{3}{4} (D-F)^2$&$0$&$0$&$0$\\
\midrule
$\alpha^{\text{(T13)}}_{KK}/\alpha^{\text{(T23)}}_{KK}$&$-\frac{1}{4} (D-F)^2$&$-\frac{1}{12} (D+3 F)^2$&$\frac{1}{12} \left(5 D^2-6 D F+9 F^2\right)$&$\frac{1}{4} (D-F)^2$&$\frac{1}{2} \left(-D^2-D F-F^2\right)$&$D F$&$\frac{1}{6} \left(D^2+3 F^2\right)$&$\frac{1}{3} D (D+3 F)$&$\frac{1}{12} \left(-5 D^2-6 D F-9 F^2\right)$&$\frac{D F}{2}$&$-D F$\\
\midrule
$\alpha^{\text{(B11)}}_{\pi\pi}$&$\frac{1}{16} (D+F)^2$&$\frac{9}{16} (D+F)^2$&$0$&$\frac{F^2}{4}$&$\frac{D^2}{4}+F^2$&$0$&$\frac{1}{16} (D-F)^2$&$\frac{9}{16} (D-F)^2$&$0$&$\frac{D^2}{4}$&$0$\\
\midrule
$\alpha^{\text{(B11)}}_{KK}$&$\frac{1}{4} (D-F)^2$&$\frac{1}{12} (D+3 F)^2$&$0$&$0$&$\frac{3}{8} (D+F)^2$&$\frac{1}{4} (D-F)^2$&$0$&$0$&$\frac{1}{12} \left(5 D^2+6 D F+9 F^2\right)$&$\frac{1}{24} (D-3 F)^2$&$\frac{1}{12} (D+3 F)^2$\\
\midrule
$\alpha^{\text{(B11)}}_{\eta\eta}$&$\frac{1}{144} (D-3 F)^2$&$\frac{1}{144} (D-3 F)^2$&$\frac{1}{36} (D-3 F)^2$&$\frac{D^2}{36}$&$\frac{D^2}{36}$&$\frac{D^2}{9}$&$\frac{1}{144} (D+3 F)^2$&$\frac{1}{144} (D+3 F)^2$&$\frac{1}{36} (D+3 F)^2$&$\frac{D^2}{36}$&$\frac{D^2}{9}$\\
\midrule
$\alpha^{\text{(B11)}}_{\pi\eta}/\alpha^{\text{(B11)}}_{\eta\pi}$&$\frac{1}{48} (D-3 F) (D+F)$&$-\frac{1}{16} (D-3 F) (D+F)$&$0$&$-\frac{D F}{12}$&$\frac{D F}{6}$&$0$&$-\frac{1}{48} (D-F) (D+3 F)$&$\frac{1}{16} (D-F) (D+3 F)$&$0$&$0$&$0$\\
\midrule
$\alpha^{\text{(B12)}}_{\pi\pi}$&$\frac{1}{3}$&$0$&$0$&$\frac{1}{48}$&$\frac{1}{12}$&$0$&$\frac{1}{48}$&$\frac{3}{16}$&$0$&$\frac{3}{16}$&$0$\\
\midrule
$\alpha^{\text{(B12)}}_{KK}$&$\frac{1}{12}$&$0$&$0$&$0$&$\frac{1}{8}$&$\frac{1}{3}$&$0$&$\frac{1}{2}$&$\frac{1}{8}$&$\frac{1}{8}$&$0$\\
\midrule
$\alpha^{\text{(B12)}}_{\eta\eta}$&$0$&$0$&$0$&$\frac{1}{48}$&$\frac{1}{48}$&$\frac{1}{12}$&$\frac{1}{48}$&$\frac{1}{48}$&$\frac{1}{12}$&$0$&$0$\\
\midrule
$\alpha^{\text{(B12)}}_{\pi\eta}/\alpha^{\text{(B12)}}_{\eta\pi}$&$0$&$0$&$0$&$-\frac{1}{48}$&$\frac{1}{24}$&$0$&$-\frac{1}{48}$&$\frac{1}{16}$&$0$&$0$&$0$\\
\midrule
$\alpha^{\text{(R11)}}_{\pi\pi}$&$\frac{5}{16} (D+F)^2$&$-\frac{3}{16} (D+F)^2$&$0$&$\frac{D^2}{6}+\frac{3 F^2}{4}$&$-\frac{D^2}{12}$&$0$&$\frac{5}{16} (D-F)^2$&$-\frac{3}{16} (D-F)^2$&$0$&$\frac{D^2}{4}$&$0$\\
\midrule
$\alpha^{\text{(R11)}}_{KK}$&$0$&$0$&$\frac{1}{12} \left(5 D^2-6 D F+9 F^2\right)$&$\frac{1}{4} (D-F)^2$&$-\frac{1}{8} (D-F)^2$&$\frac{1}{4} (D+F)^2$&$\frac{1}{6} \left(D^2+3 F^2\right)$&$\frac{1}{3} D (D+3 F)$&$0$&$\frac{1}{24} (D+3 F)^2$&$\frac{1}{12} (D-3 F)^2$\\
\midrule
$\alpha^{\text{(R11)}}_{\eta\eta}$&$\frac{1}{144} (D-3 F)^2$&$\frac{1}{144} (D-3 F)^2$&$\frac{1}{36} (D-3 F)^2$&$\frac{D^2}{36}$&$\frac{D^2}{36}$&$\frac{D^2}{9}$&$\frac{1}{144} (D+3 F)^2$&$\frac{1}{144} (D+3 F)^2$&$\frac{1}{36} (D+3 F)^2$&$\frac{D^2}{36}$&$\frac{D^2}{9}$\\
\midrule
$\alpha^{\text{(R11)}}_{\pi\eta}/\alpha^{\text{(R11)}}_{\eta\pi}$&$\frac{1}{48} (D-3 F) (D+F)$&$-\frac{1}{16} (D-3 F) (D+F)$&$0$&$-\frac{D F}{12}$&$\frac{D F}{6}$&$0$&$-\frac{1}{48} (D-F) (D+3 F)$&$\frac{1}{16} (D-F) (D+3 F)$&$0$&$0$&$0$\\
\midrule
$\alpha^{\text{(R12)}}_{\pi\pi}$&$\frac{1}{6}$&$\frac{1}{2}$&$0$&$\frac{1}{16}$&$0$&$0$&$\frac{5}{48}$&$-\frac{1}{16}$&$0$&$\frac{3}{16}$&$0$\\
\midrule
$\alpha^{\text{(R12)}}_{KK}$&$0$&$0$&$\frac{1}{8}$&$\frac{1}{12}$&$\frac{1}{3}$&$\frac{1}{12}$&$\frac{1}{24}$&$\frac{1}{8}$&$\frac{1}{4}$&$0$&$\frac{1}{4}$\\
\midrule
$\alpha^{\text{(R12)}}_{\eta\eta}$&$0$&$0$&$0$&$\frac{1}{48}$&$\frac{1}{48}$&$\frac{1}{12}$&$\frac{1}{48}$&$\frac{1}{48}$&$\frac{1}{12}$&$0$&$0$\\
\midrule
$\alpha^{\text{(R12)}}_{\pi\eta}/\alpha^{\text{(R12)}}_{\eta\pi}$&$0$&$0$&$0$&$-\frac{1}{48}$&$\frac{1}{24}$&$0$&$-\frac{1}{48}$&$\frac{1}{16}$&$0$&$0$&$0$\\
\midrule
$\alpha^{\text{(T21)}}_{\pi\pi}$&$\frac{1}{4}$&$-\frac{3}{4}$&$0$&$\frac{1}{2}$&$-1$&$0$&$\frac{1}{4}$&$-\frac{3}{4}$&$0$&$0$&$0$\\
\midrule
$\alpha^{\text{(T21)}}_{KK}$&$-\frac{1}{4}$&$-\frac{3}{4}$&$\frac{3}{4}$&$\frac{1}{4}$&$-\frac{1}{2}$&$0$&$\frac{1}{2}$&$0$&$-\frac{3}{4}$&$0$&$0$\\
\midrule
$\alpha^{\text{(T22)}}_{\pi\pi}$&$\frac{1}{4}$&$-\frac{3}{4}$&$0$&$\frac{1}{2}$&$-1$&$0$&$\frac{1}{4}$&$-\frac{3}{4}$&$0$&$0$&$0$\\
\midrule
$\alpha^{\text{(T22)}}_{KK}$&$-\frac{1}{4}$&$-\frac{3}{4}$&$\frac{3}{4}$&$\frac{1}{4}$&$-\frac{1}{2}$&$0$&$\frac{1}{2}$&$0$&$-\frac{3}{4}$&$0$&$0$\\
\midrule
$\alpha^{\text{(B21)}}_{\pi\pi}$&$\frac{1}{16} (D+F)^2$&$\frac{9}{16} (D+F)^2$&$0$&$\frac{F^2}{4}$&$\frac{D^2}{4}+F^2$&$0$&$\frac{1}{16} (D-F)^2$&$\frac{9}{16} (D-F)^2$&$0$&$\frac{D^2}{4}$&$0$\\
\bottomrule \midrule
\end{tabular}
\end{threeparttable}}%
\end{sidewaystable}

\begin{sidewaystable}
\centering
\resizebox{\textwidth}{!}{%
\begin{threeparttable}
\caption{\label{constantsTwo}Continue.}
\begin{tabular}{lcccccccccccc}
\midrule \toprule
Systems &$ND^{(*)}(I=1)$  & $ND^{(*)}(I=0)$ & $ND^{(*)}_{S}(I=\frac{1}{2})$ &$\Sigma D^{(*)}(I=\frac{3}{2})$  &$\Sigma D^{(*)}(I=\frac{1}{2})$ & $\Sigma D^{(*)}_{S}(I=1)$&$\Xi D^{(*)}(I=1)$ & $\Xi D^{(*)}(I=0)$ & $\Xi D^{(*)}_{S}(I=\frac{1}{2})$ & $\Lambda D^{(*)}(I=\frac{1}{2})$ & $\Lambda D^{(*)}_{S}(I=0)$ \\

\midrule
$\alpha^{\text{(B21)}}_{KK}$&$\frac{1}{4} (D-F)^2$&$\frac{1}{12} (D+3 F)^2$&$0$&$0$&$\frac{3}{8} (D+F)^2$&$\frac{1}{4} (D-F)^2$&$0$&$0$&$\frac{1}{12} \left(5 D^2+6 D F+9 F^2\right)$&$\frac{1}{24} (D-3 F)^2$&$\frac{1}{12} (D+3 F)^2$\\
\midrule
$\alpha^{\text{(B21)}}_{\eta\eta}$&$\frac{1}{144} (D-3 F)^2$&$\frac{1}{144} (D-3 F)^2$&$\frac{1}{36} (D-3 F)^2$&$\frac{D^2}{36}$&$\frac{D^2}{36}$&$\frac{D^2}{9}$&$\frac{1}{144} (D+3 F)^2$&$\frac{1}{144} (D+3 F)^2$&$\frac{1}{36} (D+3 F)^2$&$\frac{D^2}{36}$&$\frac{D^2}{9}$\\
\midrule
$\alpha^{\text{(B21)}}_{\pi\eta}/\alpha^{\text{(B21)}}_{\eta\pi}$&$\frac{1}{48} (D-3 F) (D+F)$&$-\frac{1}{16} (D-3 F) (D+F)$&$0$&$-\frac{D F}{12}$&$\frac{D F}{6}$&$0$&$-\frac{1}{48} (D-F) (D+3 F)$&$\frac{1}{16} (D-F) (D+3 F)$&$0$&$0$&$0$\\
\midrule
$\alpha^{\text{(B22)}}_{\pi\pi}$&$\frac{1}{16} (D+F)^2$&$\frac{9}{16} (D+F)^2$&$0$&$\frac{F^2}{4}$&$\frac{D^2}{4}+F^2$&$0$&$\frac{1}{16} (D-F)^2$&$\frac{9}{16} (D-F)^2$&$0$&$\frac{D^2}{4}$&$0$\\
\midrule
$\alpha^{\text{(B22)}}_{KK}$&$\frac{1}{4} (D-F)^2$&$\frac{1}{12} (D+3 F)^2$&$0$&$0$&$\frac{3}{8} (D+F)^2$&$\frac{1}{4} (D-F)^2$&$0$&$0$&$\frac{1}{12} \left(5 D^2+6 D F+9 F^2\right)$&$\frac{1}{24} (D-3 F)^2$&$\frac{1}{12} (D+3 F)^2$\\
\midrule
$\alpha^{\text{(B22)}}_{\eta\eta}$&$\frac{1}{144} (D-3 F)^2$&$\frac{1}{144} (D-3 F)^2$&$\frac{1}{36} (D-3 F)^2$&$\frac{D^2}{36}$&$\frac{D^2}{36}$&$\frac{D^2}{9}$&$\frac{1}{144} (D+3 F)^2$&$\frac{1}{144} (D+3 F)^2$&$\frac{1}{36} (D+3 F)^2$&$\frac{D^2}{36}$&$\frac{D^2}{9}$\\
\midrule
$\alpha^{\text{(B22)}}_{\pi\eta}/\alpha^{\text{(B22)}}_{\eta\pi}$&$\frac{1}{48} (D-3 F) (D+F)$&$-\frac{1}{16} (D-3 F) (D+F)$&$0$&$-\frac{D F}{12}$&$\frac{D F}{6}$&$0$&$-\frac{1}{48} (D-F) (D+3 F)$&$\frac{1}{16} (D-F) (D+3 F)$&$0$&$0$&$0$\\
\midrule
$\alpha^{\text{(B23)}}_{\pi\pi}$&$\frac{1}{3}$&$0$&$0$&$\frac{1}{48}$&$\frac{1}{12}$&$0$&$\frac{1}{48}$&$\frac{3}{16}$&$0$&$\frac{3}{16}$&$0$\\
\midrule
$\alpha^{\text{(B23)}}_{KK}$&$\frac{1}{12}$&$0$&$0$&$0$&$\frac{1}{8}$&$\frac{1}{3}$&$0$&$\frac{1}{2}$&$\frac{1}{8}$&$\frac{1}{8}$&$0$\\
\midrule
$\alpha^{\text{(B23)}}_{\eta\eta}$&$0$&$0$&$0$&$\frac{1}{48}$&$\frac{1}{48}$&$\frac{1}{12}$&$\frac{1}{48}$&$\frac{1}{48}$&$\frac{1}{12}$&$0$&$0$\\
\midrule
$\alpha^{\text{(B23)}}_{\pi\eta}/\alpha^{\text{(B23)}}_{\eta\pi}$&$0$&$0$&$0$&$\frac{1}{48}$&$-\frac{1}{24}$&$0$&$\frac{1}{48}$&$-\frac{1}{16}$&$0$&$0$&$0$\\
\midrule
$\alpha^{\text{(B24)}}_{\pi\pi}$&$\frac{1}{3}$&$0$&$0$&$\frac{1}{48}$&$\frac{1}{12}$&$0$&$\frac{1}{48}$&$\frac{3}{16}$&$0$&$\frac{3}{16}$&$0$\\
\midrule
$\alpha^{\text{(B24)}}_{KK}$&$\frac{1}{12}$&$0$&$0$&$0$&$\frac{1}{8}$&$\frac{1}{3}$&$0$&$\frac{1}{2}$&$\frac{1}{8}$&$\frac{1}{8}$&$0$\\
\midrule
$\alpha^{\text{(B24)}}_{\eta\eta}$&$0$&$0$&$0$&$\frac{1}{48}$&$\frac{1}{48}$&$\frac{1}{12}$&$\frac{1}{48}$&$\frac{1}{48}$&$\frac{1}{12}$&$0$&$0$\\
\midrule
$\alpha^{\text{(B24)}}_{\pi\eta}/\alpha^{\text{(B24)}}_{\eta\pi}$&$0$&$0$&$0$&$-\frac{1}{48}$&$\frac{1}{24}$&$0$&$-\frac{1}{48}$&$\frac{1}{16}$&$0$&$0$&$0$\\
\midrule
$\alpha^{\text{(R21)}}_{\pi\pi}$&$\frac{5}{16} (D+F)^2$&$-\frac{3}{16} (D+F)^2$&$0$&$\frac{D^2}{6}+\frac{3 F^2}{4}$&$-\frac{D^2}{12}$&$0$&$\frac{5}{16} (D-F)^2$&$-\frac{3}{16} (D-F)^2$&$0$&$\frac{D^2}{4}$&$0$\\
\midrule
$\alpha^{\text{(R21)}}_{KK}$&$0$&$0$&$\frac{1}{12} \left(5 D^2-6 D F+9 F^2\right)$&$\frac{1}{4} (D-F)^2$&$-\frac{1}{8} (D-F)^2$&$\frac{1}{4} (D+F)^2$&$\frac{1}{6} \left(D^2+3 F^2\right)$&$\frac{1}{3} D (D+3 F)$&$0$&$\frac{1}{24} (D+3 F)^2$&$\frac{1}{12} (D-3 F)^2$\\
\midrule
$\alpha^{\text{(R21)}}_{\eta\eta}$&$\frac{1}{144} (D-3 F)^2$&$\frac{1}{144} (D-3 F)^2$&$\frac{1}{36} (D-3 F)^2$&$\frac{D^2}{36}$&$\frac{D^2}{36}$&$\frac{D^2}{9}$&$\frac{1}{144} (D+3 F)^2$&$\frac{1}{144} (D+3 F)^2$&$\frac{1}{36} (D+3 F)^2$&$\frac{D^2}{36}$&$\frac{D^2}{9}$\\
\midrule
$\alpha^{\text{(R21)}}_{\pi\eta}/\alpha^{\text{(R21)}}_{\eta\pi}$&$\frac{1}{48} (D-3 F) (D+F)$&$-\frac{1}{16} (D-3 F) (D+F)$&$0$&$-\frac{D F}{12}$&$\frac{D F}{6}$&$0$&$-\frac{1}{48} (D-F) (D+3 F)$&$\frac{1}{16} (D-F) (D+3 F)$&$0$&$0$&$0$\\
\midrule
$\alpha^{\text{(R22)}}_{\pi\pi}$&$\frac{5}{16} (D+F)^2$&$-\frac{3}{16} (D+F)^2$&$0$&$\frac{D^2}{6}+\frac{3 F^2}{4}$&$-\frac{D^2}{12}$&$0$&$\frac{5}{16} (D-F)^2$&$-\frac{3}{16} (D-F)^2$&$0$&$\frac{D^2}{4}$&$0$\\
\midrule
$\alpha^{\text{(R22)}}_{KK}$&$0$&$0$&$\frac{1}{12} \left(5 D^2-6 D F+9 F^2\right)$&$\frac{1}{4} (D-F)^2$&$-\frac{1}{8} (D-F)^2$&$\frac{1}{4} (D+F)^2$&$\frac{1}{6} \left(D^2+3 F^2\right)$&$\frac{1}{3} D (D+3 F)$&$0$&$\frac{1}{24} (D+3 F)^2$&$\frac{1}{12} (D-3 F)^2$\\
\midrule
$\alpha^{\text{(R22)}}_{\eta\eta}$&$\frac{1}{144} (D-3 F)^2$&$\frac{1}{144} (D-3 F)^2$&$\frac{1}{36} (D-3 F)^2$&$\frac{D^2}{36}$&$\frac{D^2}{36}$&$\frac{D^2}{9}$&$\frac{1}{144} (D+3 F)^2$&$\frac{1}{144} (D+3 F)^2$&$\frac{1}{36} (D+3 F)^2$&$\frac{D^2}{36}$&$\frac{D^2}{9}$\\
\midrule
$\alpha^{\text{(R22)}}_{\pi\eta}/\alpha^{\text{(R22)}}_{\eta\pi}$&$\frac{1}{48} (D-3 F) (D+F)$&$-\frac{1}{16} (D-3 F) (D+F)$&$0$&$-\frac{D F}{12}$&$\frac{D F}{6}$&$0$&$-\frac{1}{48} (D-F) (D+3 F)$&$\frac{1}{16} (D-F) (D+3 F)$&$0$&$0$&$0$\\
\midrule
$\alpha^{\text{(R23)}}_{\pi\pi}$&$\frac{1}{6}$&$\frac{1}{2}$&$0$&$\frac{1}{16}$&$0$&$0$&$\frac{5}{48}$&$-\frac{1}{16}$&$0$&$\frac{3}{16}$&$0$\\
\midrule
$\alpha^{\text{(R23)}}_{KK}$&$0$&$0$&$\frac{1}{8}$&$\frac{1}{12}$&$\frac{1}{3}$&$\frac{1}{12}$&$\frac{1}{24}$&$\frac{1}{8}$&$\frac{1}{4}$&$0$&$\frac{1}{4}$\\
\midrule
$\alpha^{\text{(R23)}}_{\eta\eta}$&$0$&$0$&$0$&$\frac{1}{48}$&$\frac{1}{48}$&$\frac{1}{12}$&$\frac{1}{48}$&$\frac{1}{48}$&$\frac{1}{12}$&$0$&$0$\\
\midrule
$\alpha^{\text{(R23)}}_{\pi\eta}/\alpha^{\text{(R23)}}_{\eta\pi}$&$0$&$0$&$0$&$\frac{1}{48}$&$-\frac{1}{24}$&$0$&$\frac{1}{48}$&$-\frac{1}{16}$&$0$&$0$&$0$\\
\midrule
$\alpha^{\text{(R24)}}_{\pi\pi}$&$\frac{1}{6}$&$\frac{1}{2}$&$0$&$\frac{1}{16}$&$0$&$0$&$\frac{5}{48}$&$-\frac{1}{16}$&$0$&$\frac{3}{16}$&$0$\\
\midrule
$\alpha^{\text{(R24)}}_{KK}$&$0$&$0$&$\frac{1}{8}$&$\frac{1}{12}$&$\frac{1}{3}$&$\frac{1}{12}$&$\frac{1}{24}$&$\frac{1}{8}$&$\frac{1}{4}$&$0$&$\frac{1}{4}$\\
\midrule
$\alpha^{\text{(R24)}}_{\eta\eta}$&$0$&$0$&$0$&$\frac{1}{48}$&$\frac{1}{48}$&$\frac{1}{12}$&$\frac{1}{48}$&$\frac{1}{48}$&$\frac{1}{12}$&$0$&$0$\\
\midrule
$\alpha^{\text{(R24)}}_{\pi\eta}/\alpha^{\text{(R24)}}_{\eta\pi}$&$0$&$0$&$0$&$-\frac{1}{48}$&$\frac{1}{24}$&$0$&$-\frac{1}{48}$&$\frac{1}{16}$&$0$&$0$&$0$\\
\bottomrule \midrule
\end{tabular}
\end{threeparttable}}%
\end{sidewaystable}

\clearpage
\section{Impact of the decay constant}
\label{AppDecayC}
\begin{figure}[!ht]
\centering
\includegraphics[height=4.1cm,width=12.5cm]{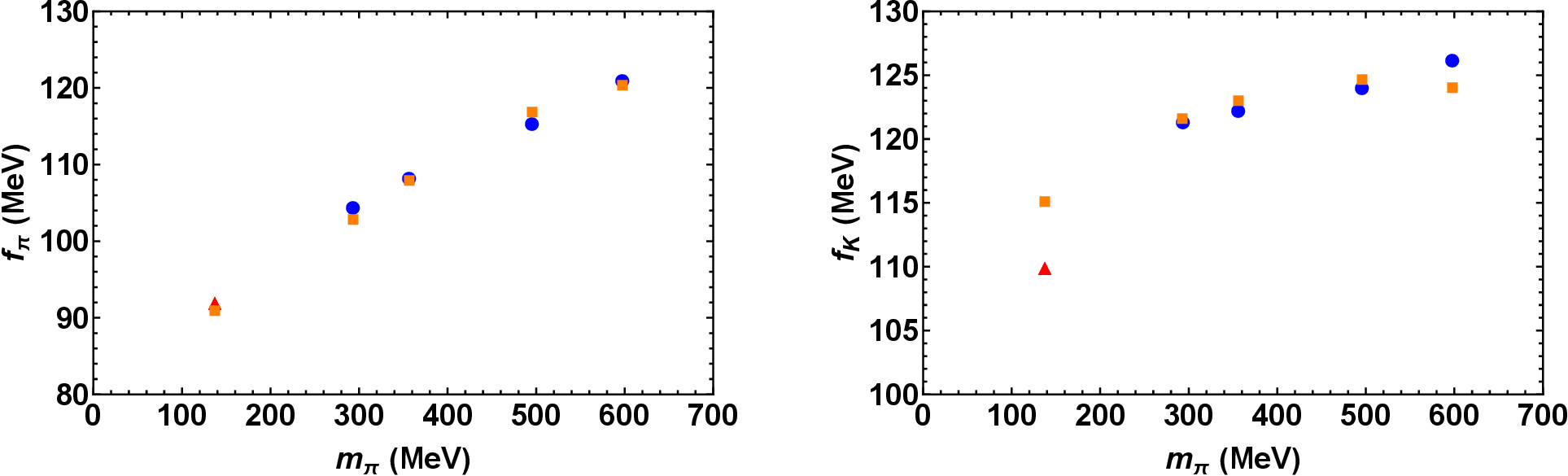}
\caption{\label{fig:fpifK}Comparison of the results of the decay constants fits through the formula, Eq.~(\ref{fpifkfeta}), to the lattice data from ref.~\cite{Walker-Loud:2008rui} and the physical values from PDG \cite{ParticleDataGroup:2022pth}. Blue circles represent the M007-M030 lattice ensembles data, red triangles indicate the physical values, and orange squares depict our findings. The errors in the data are so minimal that they are not explicitly shown in the figures.}
\end{figure}

In the main text, we use directly the physical values for the pseudoscalar-meson decay constants rather than relying on renormalization formulas, specifically Eq.~(\ref{fpifkfeta}), or the value in the chiral limit. There may, however, be differences in the outcomes when it comes to the renormalization of pseudoscalar meson decay constants or the chiral limit value. In this appendix, we explore the impact of utilizing decay constants derived from renormalization formulas or the chiral limit value on the analysis of bound states. Initially, it is essential to determine the values of $f$, $L^r_4$, and $L^r_5$ as outlined in Eq.~(\ref{fpifkfeta}). To achieve this, we fit these constants to the lattice QCD data provided in ref.~\cite{Walker-Loud:2008rui}. The fitted constants are as follows:
\begin{align}
\label{fL4L5}
f=&75.52\,\text{MeV},\nonumber\\
L^r_4=&-0.00002746,\nonumber\\
L^r_5=&0.001386,
\end{align}
with $\lambda=m_\rho=775.5\,\text{MeV}$ being assumed. The values for $f_\pi$ and $f_K$ derived from this approach are illustrated in Fig.~\ref{fig:fpifK}. Our result for $f_\pi$ closely matches the PDG value at the physical pion mass, though a slight discrepancy is noted for $f_K$. The decay constants at the physical pion mass, as determined in our chiral extrapolation, are exactly given by:
\begin{align}
\label{fpifKfetaNew}
f_\pi=&90.86\,\text{MeV},\nonumber\\
f_K=&115.03\,\text{MeV},\nonumber\\
f_\eta=&126.03\,\text{MeV}.
\end{align}
The physical values for $f_{(\pi,K,\eta)}$ are $(92.07,110.07,119.69)\,\text{MeV}$. Therefore, while our results deviate from those published by the PDG, the differences are not substantial.

Next, we calculate the bound states using the decay constants obtained from the renormalization formulas and the chiral limit. The findings are detailed in Table~\ref{tab:bdstatesDecayConstants}. The binding energies calculated using the decay constants from the renormalization formulas are nearly identical to those derived from the physical values across all channels, with the exception of the $[D\Xi]^{I=0}_{J=1/2}$ channel. However, the selection of decay constants does not influence the formation of the bound state in this channel. The binding energies calculated using the decay constants in the chiral limit exhibit minor fluctuations. Nevertheless, all bound states persist robustly even in the face of varying decay constants. Therefore, directly utilizing the physical values for the decay constants is deemed reasonable and does not impact our results.

\begin{table*}[!ht]
\centering
\resizebox{\textwidth}{!}{
\begin{threeparttable}
\caption{\label{tab:bdstatesDecayConstants}The analysis of bound states is conducted by comparing decay constants derived from physical values (PV) with those obtained through renormalization formulas (RF) and chiral limit (CL). The values of $\Delta E$ are presented in appropriate units. These states (1-10) are associated with the following channels: $([DN]^{I=0}_{J=1/2}$, $[D^{*}N]^{I=0}_{J=1/2}$, $[D^{*}N]^{I=0}_{J=3/2}$, $[D\Sigma]^{I=1/2}_{J=1/2}$, $[D^{*}\Sigma]^{I=1/2}_{J=1/2}$, $[D^{*}\Sigma]^{I=1/2}_{J=3/2}$, $[D\Xi]^{I=0}_{J=1/2}$, $[D_s\Xi]^{I=1/2}_{J=1/2}$, $[D_{s}^{*}\Xi]^{I=1/2}_{J=1/2}$, $[D_{s}^{*}\Xi]^{I=1/2}_{J=3/2})$. The first uncertainties are estimated from the LO LECs using the standard error propagation formula, while the second uncertainties are from the NLO LECs. }
\begin{tabular}{ccccccccccccccccccc}
\midrule \toprule
 decay constants & $1$ & $2$ & $3$ & $4$ & $5$ &  \\
\midrule
PV &$-13.4^{+8.8}_{-8.1}\pm 3.2$ &$-4.5^{+4.5}_{-4.6}\pm 2.1$ & $-5.0^{+5.0}_{-19.1}\pm 2.5$  &$-4.3^{+4.3}_{-4.9}\pm 2.3$&$-6.8^{+6.8}_{-3.5}\pm 2.7$ &\\
\midrule
RF &$-13.0^{+8.6}_{-8.0}\pm 3.2$&$-4.6^{+4.6}_{-4.4}\pm 2.0$ &$-5.4^{+5.4}_{-19.4}\pm 2.5$ & $-4.0^{+4.0}_{-4.7}\pm 2.3 $ & $-6.5^{+6.5}_{-3.3}\pm 2.5 $ & \\
\midrule
CL &$-24.6^{+10.7}_{-9.7}\pm 4.0 $ & $-0.2^{+0.2}_{-7.6}\pm 0.6$ & $-0.5^{+0.5}_{-13.4}\pm 1.3$ & $-10.6^{+10.6}_{-14.1}\pm 3.1$ & $-5.1^{+5.1}_{-4.0}\pm 2.2  $ & \\
\midrule
& $6$ &  $7$&  $8$ & $9$& $10$ & \\
\midrule
PV &$-4.9^{+4.9}_{-14.1} \pm 2.5 $ &$-4.1^{+0.0}_{-0.0}\pm 2.5$&$-7.9^{+3.8}_{-3.4}\pm 3.0$ &$-3.2^{+2.9}_{-0.7}\pm 2.1$ & $-0.2^{+0.2}_{-3.6}\pm 0.9$ &\\
\midrule
RF  & $-4.6^{+4.6}_{-13.9}\pm 2.5 $ & $-1.5^{+0.0}_{-0.0}\pm 1.7 $ & $-5.5^{+3.6}_{-3.2}\pm 2.6$ & $-3.2^{+2.9}_{-0.7}\pm 2.8 $ & $-0.1^{+0.1}_{-3.4}\pm 0.8 $ &\\
\midrule
CL & $-9.2^{+9.2}_{-16.3}\pm 3.1 $& $-10.7^{+0.0}_{-0.0}\pm 1.3 $ & $-10.5^{+8.2}_{-8.2}\pm 2.3$ & $-4.1^{+3.9}_{-0.9}\pm 2.4$ & $-1.3^{+1.3}_{-6.3}\pm 1.8 $ \\
\bottomrule \midrule
\end{tabular}
\end{threeparttable}}
\end{table*}

\clearpage
\section{Impact of the regulator function}
\label{AppRegulator}
In this appendix, we examine the influence of the regulator function on our results. Currently, the dependence on $\mathit{\Lambda}$ cannot be reliably  studied due to a lack of trustworthy data, such as that from experiments or lattice QCD. The current calculations are more EFT-inspired rather than being a complete EFT analysis. However, we can vary the value of $\mathit{\Lambda}$ to investigate how the bound states change, in order to observe the sensitivity of the bound states to changes in $\mathit{\Lambda}$, thereby providing a reference for future studies on the dependence of $\mathit{\Lambda}$. In addition, it is also worth investigating the impact of using different regulator functions on the results of the bound states.

Initially, we explore the influence of the cutoff parameter $\mathit{\Lambda}$ on the bound states. As the value of $\mathit{\Lambda}$ increases, we first determine $c_s$ by examining the bound states in the $[DN]^{I=0}_{J=1/2}$, $[D\Sigma]^{I=1/2}_{J=1/2}$, and $[D_s\Xi]^{I=1/2}_{J=1/2}$ channels, which do not involve $c_t$. We then determine $c_t$ based on the bound states in the $([D^{*}N]^{I=0}_{J=1/2}$, $[D^{*}N]^{I=0}_{J=3/2}$, $[D^{*}\Sigma]^{I=1/2}_{J=1/2}$, $[D^{*}\Sigma]^{I=1/2}_{J=3/2}$, $[D_{s}^{*}\Xi]^{I=1/2}_{J=1/2}$, $[D_{s}^{*}\Xi]^{I=1/2}_{J=3/2})$ channels. For $C_{\text{NLO}}$, we also adjust its range with increasing $\mathit{\Lambda}$ to ensure a reasonable uncertainty in the binding energy. For $\mathit{\Lambda}=(500/600)$ MeV, we set $(-60/-40) \,\text{GeV}^{-4}<c_{\text{NLO}}<(60/40)\,\text{GeV}^{-4}$. However, for the $[D\Xi]^{I=0}_{J=1/2}$ chanel, we keep $-100 \,\text{GeV}^{-4}<c_{\text{NLO}}<-100\,\text{GeV}^{-4}$ due to the absence of LO LECs, which results in a larger uncertainty from the NLO LECs. Note that, the cutoff parameter $\mathit{\Lambda}$ and the coupling constants cannot vary independently of each other because of the mutual correlations, i.e., $c_{(s,t,\text{NLO})}=c_{(s,t,\text{NLO})}(\mathit{\Lambda})$. Subsequently, the summarized findings are presented in Table~\ref{tab:bdstatesLambda}.

It is evident that all bound states persist within the range of $\mathit{\Lambda}$ from 400 to 600 MeV. Channel 7 exhibits a deepening trend with increasing $\mathit{\Lambda}$, while all other channels remain unchanged within errors. We can see that channel 7 does not have LO contact contribution. Therefore, we can only achieve a larger uncertainty in the binding energy when the NLO LECs do not decrease with increasing $\mathit{\Lambda}$. Additional contact terms are required for a comprehensive description in this channel, leading to results independent of the cutoff $\mathit{\Lambda}$. Higher-order calculations will be pursued as more lattice-QCD or experimental data become available to determine the LECs.

Moving on, we investigate the impact of the parameter $n$ on the bound states, and the results are detailed in Table~\ref{tab:bdstatesn}. The bound states generally deepen as $n$ increases, yet the outcomes exhibit a clear convergence tendency. This convergence suggests that higher-order calculations will yield robust results.

The different functional form for the regulators may lead to very different results. In this investigation, we delve into the bound states using a regulator function in the monopole form:
\begin{align}
\label{eqmonopole}
\mathcal{F}(\bm{q}) =\frac{\mathit{\Lambda}^{2n}}{\bm{q}^{2n}+\mathit{\Lambda}^{2n}}.
\end{align}
The results also show minor fluctuations, as depicted in Table~\ref{tab:bdstatesRF}. Nonetheless, all bound states persist in the different regulator functions and remain consistent within errors. In addition, mirroring the approach taken with the Gaussian-form, we adjust $\mathit{\Lambda}$ and $c_{s,t}$. This adjustment allows us to examine the dynamics of all bound states, with our observations compiled in Table~\ref{tab:bdstatesLambdaNew}. All bound states persist within the $\mathit{\Lambda}$ range of 400 to 600 MeV. Furthermore, we explore the influence of the parameter $n$ on bound states when employing the monopole-form regulator function, as detailed in Table~\ref{tab:bdstatesnNew}. It becomes evident that this approach yields better convergence compared to the Gaussian-form case.

In conclusion, our detailed examination sheds light on the role of the regulator function. We find that the persistence of all bound states across various regulator functions suggests a possibility: the results may be independent of the choice of regulator function. Nonetheless, to conclusively affirm this finding, further high-order calculations with reliable data are needed.
\begin{table*}[!ht]
\centering
\resizebox{\textwidth}{!}{
\begin{threeparttable}
\caption{\label{tab:bdstatesLambda} The bound states are analyzed concerning the variations in $(\mathit{\Lambda},c_s, c_t)$, with $n=3$ within the SU(3) frameworks, utilizing a Gaussian-form regulator function. The values of $(\mathit{\Lambda},c_s, c_t)$ and $\Delta E$ are presented in appropriate units. These states (1-10) are associated with the following channels: $([DN]^{I=0}_{J=1/2}$, $[D^{*}N]^{I=0}_{J=1/2}$, $[D^{*}N]^{I=0}_{J=3/2}$, $[D\Sigma]^{I=1/2}_{J=1/2}$, $[D^{*}\Sigma]^{I=1/2}_{J=1/2}$, $[D^{*}\Sigma]^{I=1/2}_{J=3/2}$, $[D\Xi]^{I=0}_{J=1/2}$, $[D_s\Xi]^{I=1/2}_{J=1/2}$, $[D_{s}^{*}\Xi]^{I=1/2}_{J=1/2}$, $[D_{s}^{*}\Xi]^{I=1/2}_{J=3/2})$. The first uncertainties are estimated from the LO LECs using the standard error propagation formula, while the second uncertainties are from the NLO LECs. }
\begin{tabular}{ccccccccccccccccccc}
\midrule \toprule
$(\mathit{\Lambda},c_s,c_t)$  & $1$ & $2$ & $3$ & $4$ & $5$ \\
\midrule
$(400,-8.2^{+3.8}_{-3.4},1.1^{+1.5}_{-2.2})$ &$-13.4^{+8.8}_{-8.1}\pm 3.2$ &$-4.5^{+4.5}_{-4.6}\pm 2.1$ & $-5.0^{+5.0}_{-19.1}\pm 2.5$  &$-4.3^{+4.3}_{-4.9}\pm 2.3$&$-6.8^{+6.8}_{-3.5}\pm 2.7$ & \\
\midrule
$(500,-5.6^{+3.4}_{-3.2},1.0^{+1.7}_{-2.1})$ &$-13.2^{+9.2}_{-7.9}\pm 3.8$&$-5.9^{+5.9}_{-5.2}\pm 3.5$ &$-6.8^{+6.8}_{-22.3}\pm3.9$ & $-4.2^{+4.2}_{-4.2}\pm 2.5 $ & $-3.4^{+3.4}_{-7.1}\pm 4.6$ & \\
\midrule
$(600,-3.9^{+3.2}_{-2.9},0.9^{+1.4}_{-1.9})$ &$-12.2^{+8.3}_{-6.1}\pm 4.9$ &$-6.6^{+6.6}_{-6.3}\pm 4.8$ &$-8.5^{+8.5}_{-17.2}\pm 5.2 $ & $-4.5^{+4.5}_{-4.9}\pm 2.7$ & $-2.6^{+2.6}_{-9.1}\pm 5.7$ & \\
\midrule
  & $6$ &  $7$&  $8$ & $9$& $10$ & \\
\midrule
$(400,-8.2^{+3.8}_{-3.4},1.1^{+1.5}_{-2.2})$   &$-4.9^{+4.9}_{-14.1} \pm 2.5 $ &$-4.1^{+0.0}_{-0.0}\pm 2.5$&$-7.9^{+3.8}_{-3.4}\pm 3.0$ &$-3.2^{+2.9}_{-0.7}\pm 2.1$ & $-0.2^{+0.2}_{-3.6}\pm 0.9$ & \\
\midrule
$(500,-5.6^{+3.4}_{-3.2},1.0^{+1.7}_{-2.1})$ & $-3.8^{+3.8}_{-16.4}\pm 4.1 $ & $-24.3^{+0.0}_{-0.0}\pm 16.7$ & $-12.9^{+9.3}_{-11.4}\pm 5.9$ & $-1.8^{+1.7}_{-2.4}\pm 3.7$ & $-0.7^{+0.7}_{-5.6}\pm 1.2 $ &\\
\midrule
$(600,-3.9^{+3.2}_{-2.9},0.9^{+1.4}_{-1.9})$ & $-1.3^{+1.3}_{-21.5}\pm 7.1$ & $-45.0^{+0.0}_{-0.0}\pm 32.9$ & $-11.5^{+11.5}_{-15.2}\pm 6.7$ & $-0.6^{+0.6}_{-2.8}\pm 4.8 $ & $-0.9^{+0.9}_{-5.1}\pm 1.4$ &\\
\bottomrule \midrule
\end{tabular}
\end{threeparttable}}
\end{table*}
\begin{table*}[!ht]
\centering
\resizebox{\textwidth}{!}{
\begin{threeparttable}
\caption{\label{tab:bdstatesn} The bound states are analyzed concerning the variations in $n$ using a Gaussian-form regulator function, with $(\mathit{\Lambda}=400, c_s=-8.2^{+3.8}_{-3.4}, c_t=1.1^{+1.5}_{-2.2})$ in the SU(3) frameworks. The notation is the same as in Table~\ref{tab:bdstatesLambda}.}
\begin{tabular}{ccccccccccccccccccc}
\midrule \toprule
   & $1$ & $2$ & $3$ & $4$ & $5$ &  \\
\midrule
$n=2$ &$-11.8^{+8.3}_{-7.6}\pm 3.3$&$-3.6^{+3.6}_{-4.1}\pm 1.9$ &$-4.0^{+4.0}_{-17.9}\pm 2.4$ & $-3.4^{+3.4}_{-4.1}\pm 2.2 $ & $-5.7^{+5.7}_{-3.1}\pm 2.6$ &\\
\midrule
$n=3$ &$-13.4^{+8.8}_{-8.1}\pm 3.2$ &$-4.5^{+4.5}_{-4.6}\pm 2.1$ & $-5.0^{+5.0}_{-19.1}\pm 2.5$  &$-4.3^{+4.3}_{-4.9}\pm 2.3$&$-6.8^{+6.8}_{-3.5}\pm 2.7$ &  \\
\midrule
$n=4$ &$-14.5^{+9.6}_{-8.8}\pm 3.3 $&$-5.1^{+5.1}_{-4.9}\pm 2.2 $ &$-5.7^{+5.7}_{-21.2}\pm 2.6$ & $-4.9^{+4.9}_{-5.5}\pm 2.4$ & $-7.5^{+7.5}_{-4.4}\pm 2.8$ & \\
\midrule
  & $6$ &  $7$&  $8$ & $9$& $10$ & \\
\midrule
$n=2$  & $-4.0^{+4.0}_{-13.2}\pm 2.4$ & $-3.3^{+0.0}_{+0.0}\pm 2.4$ & $-6.8^{+3.5}_{-3.1}\pm 3.0$ & $-2.5^{+2.2}_{-0.4}\pm 1.5$ & $-0.1^{+0.1}_{-3.2}\pm 0.6$ &\\
\midrule
$n=3$ &$-4.9^{+4.9}_{-14.1} \pm 2.5 $ &$-4.1^{+0.0}_{-0.0}\pm 2.5$&$-7.9^{+3.8}_{-3.4}\pm 3.0$ &$-3.2^{+2.9}_{-0.7}\pm 2.1$ & $-0.2^{+0.2}_{-3.6}\pm 0.9$ &  \\
\midrule
$n=4$ & $-5.5^{+5.5}_{-15.9}\pm 2.6 $ & $-4.6^{+0.0}_{-0.0}\pm 2.6 $ & $-8.6^{+4.6}_{-4.8}\pm 3.1 $ & $-3.6^{+3.3}_{-1.2}\pm 2.3 $ & $-0.3^{+0.3}_{-4.3}\pm 1.1 $ &\\
\bottomrule \midrule
\end{tabular}
\end{threeparttable}}
\end{table*}
\begin{table*}[!ht]
\centering
\resizebox{\textwidth}{!}{
\begin{threeparttable}
\caption{\label{tab:bdstatesRF}The bound states are analyzed in the different regulator function: Gaussian-form (GF) and monopole-form (MF). The notation is the same as in Table~\ref{tab:bdstatesLambda}}
\begin{tabular}{ccccccccccccccccccc}
\midrule \toprule
 & $1$ & $2$ & $3$ & $4$ & $5$ & \\
\midrule
GF &$-13.4^{+8.8}_{-8.1}\pm 3.2$ &$-4.5^{+4.5}_{-4.6}\pm 2.1$ & $-5.0^{+5.0}_{-19.1}\pm 2.5$  &$-4.3^{+4.3}_{-4.9}\pm 2.3$&$-6.8^{+6.8}_{-3.5}\pm 2.7$ & \\
\midrule
MF &$-23.4^{+13.5}_{-11.3}\pm 8.4$&$-6.8^{+6.8}_{-6.2}\pm 4.6$ &$-10.1^{+10.1}_{-27.8}\pm 6.3$ & $-7.5^{+7.5}_{-8.1}\pm 5.8$ & $-10.6^{+10.6}_{-5.2}\pm 6.2$ & \\
\midrule
 & $6$ &  $7$&  $8$ & $9$& $10$ & \\
\midrule
GF &$-4.9^{+4.9}_{-14.1} \pm 2.5 $ &$-4.1^{+0.0}_{-0.0}\pm 2.5$&$-7.9^{+3.8}_{-3.4}\pm 3.0$ &$-3.2^{+2.9}_{-0.7}\pm 2.1$ & $-0.2^{+0.2}_{-3.6}\pm 0.9$ & \\
\midrule
MF & $-9.3^{+9.3}_{-19.6}\pm 6.5$ & $-8.6^{+0.0}_{-0.0}\pm 6.7$ & $-14.1^{+7.6}_{-8.3}\pm 7.8$ & $-5.8^{+5.4}_{-1.5}\pm 5.6$ & $-0.9^{+0.9}_{-8.4}\pm 3.3$ &\\
\bottomrule \midrule
\end{tabular}
\end{threeparttable}}
\end{table*}
\begin{table*}[!ht]
\centering
\resizebox{\textwidth}{!}{
\begin{threeparttable}
\caption{\label{tab:bdstatesLambdaNew}The bound states are analyzed concerning the variations in $(\mathit{\Lambda}, c_s, c_t)$, with $n=3$ within the SU(3) frameworks, utilizing a monopole-form regulator function. The notation is the same as in Table~\ref{tab:bdstatesLambda}
}
\begin{tabular}{ccccccccccccccccccc}
\midrule \toprule
$(\mathit{\Lambda},c_s,c_t)$  & $1$ & $2$ & $3$ & $4$ & $5$ \\
\midrule
$(400,-8.2^{+3.8}_{-3.4},1.1^{+1.5}_{-2.2})$ &$-23.4^{+13.5}_{-11.3}\pm 8.4$&$-6.8^{+6.8}_{-6.2}\pm 4.6$ &$-10.1^{+10.1}_{-27.8}\pm 6.3$ & $-7.5^{+7.5}_{-8.1}\pm 5.8$ & $-10.6^{+10.6}_{-5.2}\pm 6.2$ & \\
\midrule
$(500,-5.6^{+2.5}_{-3.1},1.0^{+1.7}_{-2.0})$ &$-23.2^{+12.8}_{-10.3}\pm 8.3$&$-7.9^{+7.9}_{-8.9}\pm 5.5$ &$-12.8^{+12.8}_{-26.3}\pm 7.9$ & $-7.2^{+7.2}_{-9.2}\pm 5.5 $ & $-8.4^{+8.4}_{-6.1}\pm 9.6$ & \\
\midrule
$(600,-3.9^{+2.1}_{-2.5},0.9^{+1.5}_{-1.8})$ &$-22.2^{+14.1}_{-15.1}\pm 9.4$ &$-8.6^{+8.6}_{-9.1}\pm 6.8$ &$-13.5^{+13.5}_{-27.5}\pm 5.8 $ & $-7.5^{+7.5}_{-8.7}\pm 5.7$ & $-6.6^{+6.6}_{-6.4}\pm 9.7$ & \\
\midrule
  & $6$ &  $7$&  $8$ & $9$& $10$ & \\
\midrule
$(400,-8.2^{+3.8}_{-3.4},1.1^{+1.5}_{-2.2})$   & $-9.3^{+9.3}_{-19.6}\pm 6.5$ & $-8.6^{+0.0}_{-0.0}\pm 6.7$ & $-14.1^{+7.6}_{-8.3}\pm 7.8$ & $-5.8^{+5.4}_{-1.5}\pm 5.6$ & $-0.9^{+0.9}_{-8.4}\pm 3.3$ &\\
\midrule
$(500,-5.6^{+2.5}_{-3.1},1.0^{+1.7}_{-2.0})$ & $-7.8^{+7.8}_{-18.2}\pm 5.1 $ & $-28.3^{+0.0}_{-0.0}\pm 22.7$ & $-16.9^{+14.2}_{-15.7}\pm 9.9$ & $-3.8^{+3.8}_{-4.2}\pm 5.2$ & $-1.5^{+1.5}_{-7.2}\pm 3.6 $ &\\
\midrule
$(600,-3.9^{+2.1}_{-2.5},0.9^{+1.5}_{-1.8})$ & $-5.3^{+5.3}_{-16.5}\pm 7.6$ & $-49.8^{+0.0}_{-0.0}\pm 38.6$ & $-18.5^{+16.3}_{-18.3}\pm 11.2$ & $-1.6^{+1.6}_{-3.5}\pm 4.6 $ & $-2.3^{+2.3}_{-6.4}\pm 3.4$ &\\
\bottomrule \midrule
\end{tabular}
\end{threeparttable}}
\end{table*}
\begin{table*}[!ht]
\centering
\resizebox{\textwidth}{!}{
\begin{threeparttable}
\caption{\label{tab:bdstatesnNew}The bound states are analyzed concerning the variations in $n$ using a monopole-form regulator function, with $(\mathit{\Lambda}=400, c_s=-8.2^{+3.8}_{-3.4}, c_t=1.1^{+1.5}_{-2.2})$ in the SU(3) frameworks. The notation is the same as in Table~\ref{tab:bdstatesLambda}.}
\begin{tabular}{ccccccccccccccccccc}
\midrule \toprule
   & $1$ & $2$ & $3$ & $4$ & $5$ &  \\
\midrule
$n=2$ &$-30.1^{+16.5}_{-14.8}\pm 29.9$ & $-6.6^{+6.6}_{-6.4}\pm 7.8 $ &$-12.6^{+12.6}_{-32.6}\pm 14.0$ & $-8.4^{+8.4}_{-9.2}\pm 12.1$ & $-11.2^{+11.2}_{-6.1}\pm 12.4 $ & \\
\midrule
$n=3$ &$-23.4^{+13.5}_{-11.3}\pm 8.4$&$-6.8^{+6.8}_{-6.2}\pm 4.6$ &$-10.1^{+10.1}_{-27.8}\pm 6.3$ & $-7.5^{+7.5}_{-8.1}\pm 5.8$ & $-10.6^{+10.6}_{-5.2}\pm 6.2$ & \\
\midrule
$n=4$ &$-21.6^{+14.5}_{-10.4}\pm 6.3 $&$-7.0^{+7.0}_{-6.6}\pm 3.8 $ &$-9.4^{+9.4}_{-27.9}\pm 4.9 $ & $-7.3^{+7.3}_{-7.9}\pm 4.6 $ & $-10.4^{+10.6}_{-9.1}\pm 4.9 $ & \\
\midrule
 & $6$ &  $7$&  $8$ & $9$& $10$ & \\
\midrule
$n=2$  & $-11.5^{+11.5}_{-23.2}\pm 15.2 $ & $-12.0^{+0.0}_{-0.0}\pm 11.0$ & $-18.1^{+10.7}_{-9.2}\pm 17.4 $ & $-6.6^{+6.6}_{-2.9}\pm 12.2 $ & $-1.1^{+1.1}_{-9.7}\pm 8.1 $ &\\
\midrule
$n=3$  & $-9.3^{+9.3}_{-19.6}\pm 6.5$ & $-8.6^{+0.0}_{-0.0}\pm 6.7$ & $-14.1^{+7.6}_{-8.3}\pm 7.8$ & $-5.8^{+5.4}_{-1.5}\pm 5.6$ & $-0.9^{+0.9}_{-8.4}\pm 3.3$ &\\
\midrule
$n=4$  & $-8.7^{+8.7}_{-21.9}\pm 4.0 $ & $-7.8^{+0.0}_{-0.0}\pm 5.0 $ & $-13.1^{+7.4}_{-3.9}\pm 5.8$ & $-5.6^{+5.3}_{-1.4}\pm 4.4$ & $-0.8^{+0.8}_{-8.0}\pm 2.6 $ &\\
\bottomrule \midrule
\end{tabular}
\end{threeparttable}}
\end{table*}

\clearpage
\bibliographystyle{utphys}
\bibliography{OctetBaryonandHeavyMesonInteraction}

\end{document}